\documentclass[namedreferences]{solarphysics}

\usepackage[optionalrh]{spr-sola-addons} % For Solar Physics
\usepackage{epsfig}          % For eps figures, old commands
\usepackage{graphicx,color}        % For eps figures, newer & more powerfull
\usepackage{amssymb}         % useful mathematical symbols
\usepackage{color}           % For color text: \color command
\usepackage{url}             % For breaking URLs easily trough lines
\newcommand{\etal}{{\it et al.}}

\usepackage{setspace}
\usepackage{rotating}
\usepackage{float}
\usepackage{textcomp}
\usepackage[colorlinks=true,citecolor=blue,linkcolor=red]{hyperref}
\usepackage[english]{babel}
\usepackage{graphicx}
\usepackage[mediumspace,mediumqspace,Grey,squaren]{SIunits}
\usepackage{tabularx}

\newcommand{\aap}{    {\it Astron. Astrophys.}}

\newcommand{\apj}{    {\it Astrophys. J.}}
\newcommand{\apjl}{   {\it Astrophys. J. Lett.}}

\newcommand{\jgr}{    {\it J. Geophys. Res.}}

\newcommand{\pasj}{   {\it Pub. Astron. Soc. Japan}}

\newcommand{\solphys}{{\it Solar Phys.}}

\newcommand{\ssr}{    {\it Space Sci. Rev.}}

\chardef\us=`\_
\begin{document}

\begin{article}
\begin{opening}

\title{Transverse oscillations in a coronal loop triggered by a jet}

%\author[addressref={aff1},corref,email={shilpa.sarkar30@gmail.com}]{\inits{S. Sarkar}\fnm{S. Sarkar}~\lnm{}}%\sep
%\author[addressref={aff2},email={vaibhav@iiap.res.in}]{\inits{V. Pant}\fnm{V. Pant}~\lnm{}}%\sep
%\author[addressref={aff3}]{\inits{A. K. Srivastava}\fnm{A. K. Srivastava}~\lnm{}}%\sep
%\author[addressref={aff2,aff4}]{\inits{D. Banerjee}\fnm{D. Banerjee}~\lnm{}}%\sep
%\author[addressref={aff5}]{\inits{M. Ruderman}\fnm{M. Ruderman}~\lnm{}}%\sep

%\address[id=aff1]{Presidency University, 86/1, College Street, Kolkata 700 073, West Bengal, India}
%\address[id=aff2]{Indian Institute of Astrophysics, Koramangala IInd Block, Bangalore 560 063, India}
%\address[id=aff4]{Center of Excellence in Space Sciences, IISER Kolkata, Mohanpur 741 252, West Bangal, India
%West Bengal, India}

%\runningauthor{S. Sarkar et al.}
%\runningtitle{Transverse oscillations in a coronal loop triggered by a jet}
\author{S.~\surname{Sarkar}$^{1}$\sep V.~\surname{Pant}$^{2}$\sep A. K.~\surname{Srivastava}$^{3}$\sep D.~\surname{Banerjee}$^{2,4}$}

\runningauthor{Sarkar, S. \etal}
\runningtitle{Transverse oscillations in a coronal loop triggered by a jet}
\institute{$^{1}$Aryabhatta Research Institute of Observational Sciences, Manora Peak, Nainital 263002, India. email: \url{shilpa@aries.res.in}\\
           $^{2}$Indian Institute of Astrophysics, Koramangala, Bangalore 560034, India. email: \url{vaibhav@iiap.res.in}\\
           $^{3}$Indian Institute of Technology, Varanasi 221005, India \\
           $^{4}$Center of Excellence in Space Sciences, IISER, Kolkata 741252, India.\\
                   }

\begin{abstract}
We detect and analyse transverse oscillations in a coronal loop, lying at the south east limb of the Sun as seen from the \textit{{Atmospheric Imaging Assembly}} (AIA) onboard \textit{{Solar Dynamics Observatory}} (SDO). The jet is believed to trigger transverse oscillations in the coronal loop. The jet originates from a region close to the coronal loop on 19$^{\rm th}$ September 2014 at 02:01:35 UT. The length of the loop is estimated to be { between 377-539~Mm}. Only one complete oscillation is detected with an average period of about $32\pm5$~min. Using MHD seismologic inversion techniques, we estimate the magnetic field inside the coronal loop to be { between $2.68 -4.5$~G. The velocity of {the} hot and cool {components} of the jet is estimated to be 168~{km~s$^{-1}$} and 43~{km~s$^{-1}$}, respectively}. The energy density of the jet is found to be greater than the energy density of the oscillating coronal loop. Therefore, we conclude that the jet {triggered} transverse oscillations in the coronal loop. To our knowledge, this is the first {coronal loop seismology study} using the properties of {a} jet propagation {to trigger} oscillations.
\end{abstract}

\keywords{Sun; Oscillations; Magnetic fields; Jets}
\end{opening}

%-------------------------------------------------

%=====================================================

\section{Introduction}

Magnetohydrodynamics (MHD) waves are ubiquitous in solar atmosphere. With the advent of the high resolution observations after the launch of \textit{{Solar and Heliospheric Observatory}} (SOHO), \textit{{Transition Region and Coronal Explorer}} (TRACE), \textit{Hinode}, \textit{Solar Terrestrial Relations Observatory} (STEREO) and {\textit{Solar Dynamics Observatory}} (SDO), MHD waves have been well studied in recent years \citep{afsa,nodrd,2002ApJ...580L..85O,2007A&A...473L..13O,2009ApJ...698..397V,2010NewA...15....8S,asch2011,white2012,2013SSRv..175....1M}.

The MHD waves and oscillations provide an important input in diagnosing the local plasma conditions using the principles of MHD seismology; {as} first suggested by \inlinecite{uchida1970} using Moreton wave and \inlinecite{rosenberg1970} using intensity fluctuations associated with type IV radio emission. Using the principles of MHD seismology, magnetic field strength in the solar corona {was} estimated by \inlinecite{afsa},  \inlinecite{nakariakov2000},  \inlinecite{no},  \inlinecite{schrij2002},  \inlinecite{asch2002},  \inlinecite{nv},  \inlinecite{asch2006},  \inlinecite{ruderman2009},  \inlinecite{asch2011}.

 \inlinecite{er}, \inlinecite{roberts1984} interpreted transverse oscillations as nearly incompressible fast kink mode in {the} MHD regime. The first {observations} of transverse oscillations in coronal {loops were} reported by \inlinecite{nodrd}, \inlinecite{afsa}, \inlinecite{schrij1999}.  Transverse loop oscillations are often excited by nearby flares \citep{afsa,nodrd,hori,verwichte2010,wang2012} and reconnection at loop top \citep{white2012}.
 %he main mechanism of MHD seismology is to combine the properties of magnetohydrodynamic waves and oscillations-periods, wavelengths, amplitudes, temporal and spatial signatures as well as some physical plasma parameters of the medium like temperature, density etc.measured from the data available (the observations), with a given theoretical model of wave phenoma (dispersion relations etc.). 
%This enables us to determine the mean parameters of the corona, such as magnetic field strength \citep{afsa, nakariakov2000, no, schrij2002, asch2002, asch2004, nv, asch2006, ruderman2009, asch2011} and resistivity and viscosity \citep{nodrd}. 
Other methods of measuring coronal magnetic field include gyro-resonance modeling of radio emission \citep{lee1999}, but {this} can only be applied to strong magnetic field regions like sunspots.

Transverse oscillations are often found to be damped most likely due to resonant absorption \citep{ruderman2002,gaa,holweg88}. Recently there have been few reports on decayless oscillations in coronal loops \citep{niscito13,anfi}. The driving mechanisms of such oscillations have not been clearly understood. Recently, \inlinecite{zimovets15} have provided a statistical investigation of coronal loop oscillations observed with SDO in association with blast waves due to a nearby flare, coronal mass ejections, type II radio bursts, etc. They found that {kink} oscillations ($\sim$ 95 $\%$ of them) were triggered by {nearby} low coronal eruptions (LCE) observed in the extreme ultraviolet band. Thus different {types of transient} can trigger oscillations in nearby magnetic structures which provide additional data for coronal seismology. Here we explore if transients like jets, that carry much smaller energy as compared to CMEs or blast {waves can} trigger oscillations in nearby coronal loops.  

In {Section}~\ref{obs}, we report the observations and the data analysis. In {Section}~\ref{stereo}, we describe the observations from STEREO. In {Section}~\ref{sec4} we present time distance analysis to study the dynamics of the jet which is followed by {Section}~\ref{mhd}, where we carry out MHD seismology to estimate the magnetic field strength. In {Section}~\ref{energy}, we calculate the energy stored in jet and coronal loop oscillations. In {Section}~\ref{sec7}, we report the coupling between oscillations in jet with coronal loop oscillations which is followed by conclusions.

%In this work, we report long period, decayless coronal loop oscillations excited by nearby oscillating jet.

\section{Observations and Data analysis}\label{obs}
\begin{figure}[h!]
  \begin{center}
   \includegraphics[width=11cm, height=10cm]{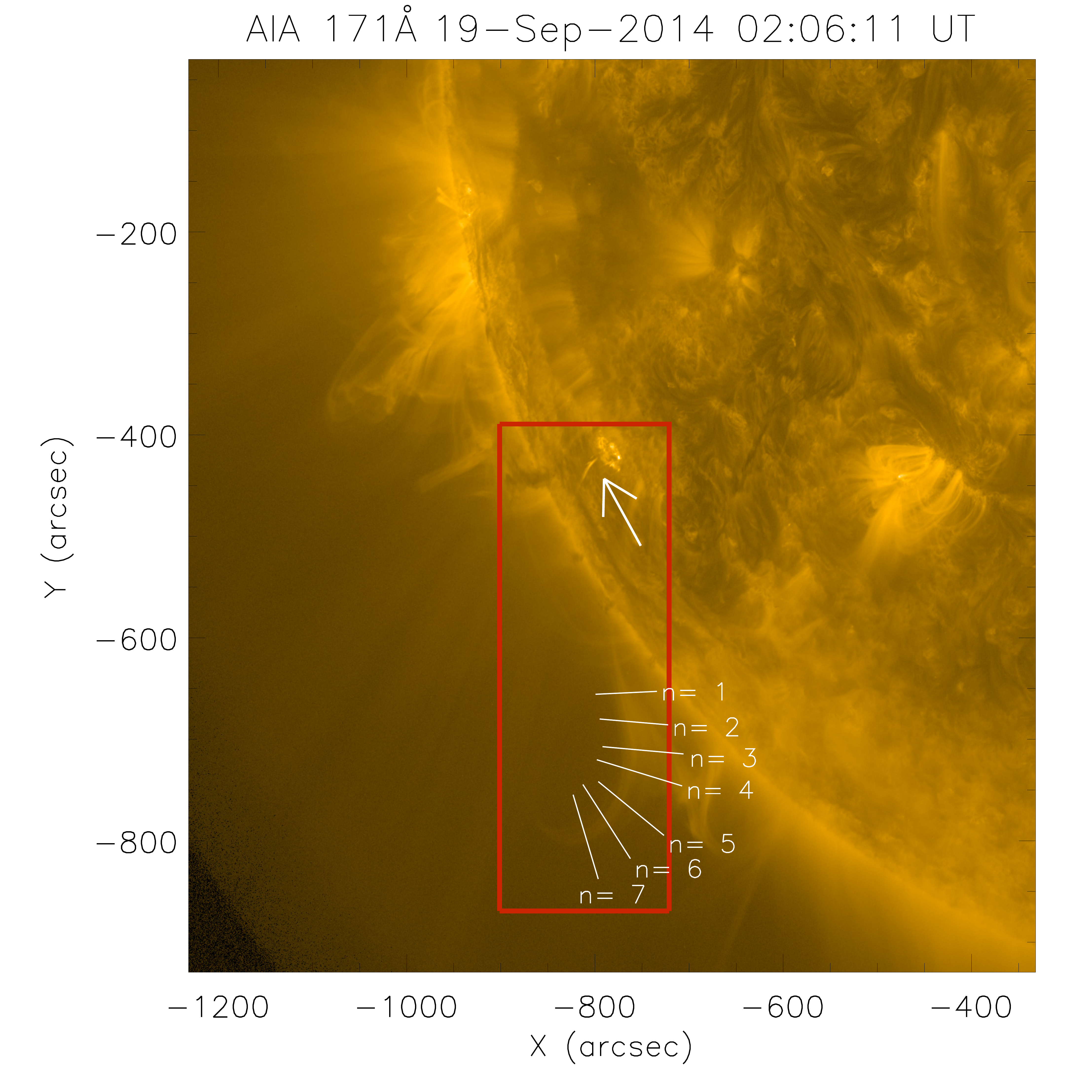}
    \caption{AIA/SDO image of 171~\AA, taken on 19${\rm th}$ September 2014 at 02:06:11~UT. The red box highlights the jet (top) and the coronal loop (bottom). {The white} arrow points {at the} location {of the} jet. Seven artificial slices are placed perpendicular to different parts of the loop to detect the transverse oscillations ({see Figure}~\ref{r4}). An animation is available online as movie~1.}
    \label{r2}
  \end{center}
\end{figure}

A jet was observed at the south east limb of the Sun on 19$^{\rm th}$ September 2014 at 02:01:35~UT. A narrow CME was detected at 04:37 UT in {\textit{Large Angle Spectroscopic Coronagraph}} (LASCO) on board SOHO. The CME may be the coronographic counterpart of the jet \citep{feng2012,paraschiv2010,nisticobothmer2009}. The observation was made using the {extreme ultraviolet} (EUV) passbands of {the \textit{Atmospheric Imaging Assembly}} (AIA) onboard SDO. AIA instrument provides almost simultaneous full-disk images of the Sun {at} ten different wavelengths, of which seven are in the EUV band. AIA has a spatial resolution of $1.3''$, pixel size of $0.6''$ and a cadence of 12~s \citep{lemen2011}. At 02:12:11~UT ($\sim$ 12 min after the jet was started), transverse oscillations in a coronal loop were observed. The region where the jet and the coronal loop were observed is outlined with a red box in {Figure}~\ref{r2}. The jet is marked with an arrow. A sequence of images was taken, encompassing 1.5~hr, covering 1~min before and 1.48~hr after the jet. Since the jet gets fainter as it propagates outward, we make an unsharp mask movie to clearly show the propagation of the jet and its interaction with the coronal loop ({see movie~1} online). We note that only one complete transverse oscillation of coronal loop was clearly observed. Moreover we also note that the jet oscillates in the transverse direction to its propagation ({see Section}~\ref{sec4}).

%%%%%%%%%%%%%%%%%%%%%%%%%%%%%%%%%%%%%%%%%%%%%
%%%%%%%%%%%%%%%%%%%%%%%%%%%%%%%%%%%%%%%%%%%%%
%%%%%%%%%%%%%%%%%%%%%%%%%%%%%%%%%%%%%%%%%%%%%
%%%%%%%%%%%%%%%%%%%%%%%%%%%%%%%%%%%%%%%%%%%%%

%%%%%%%%%%%%%%%%%%%%%%%%%%%%%%%%%%%%%%%%%%%%%
%%%%%%%%%%%%%%%%%%%%%%%%%%%%%%%%%%%%%%%%%%%%%
\begin{figure} 
\centerline{\hspace*{0.015\textwidth}
               \includegraphics[width=0.515\textwidth,height=10cm,clip=]{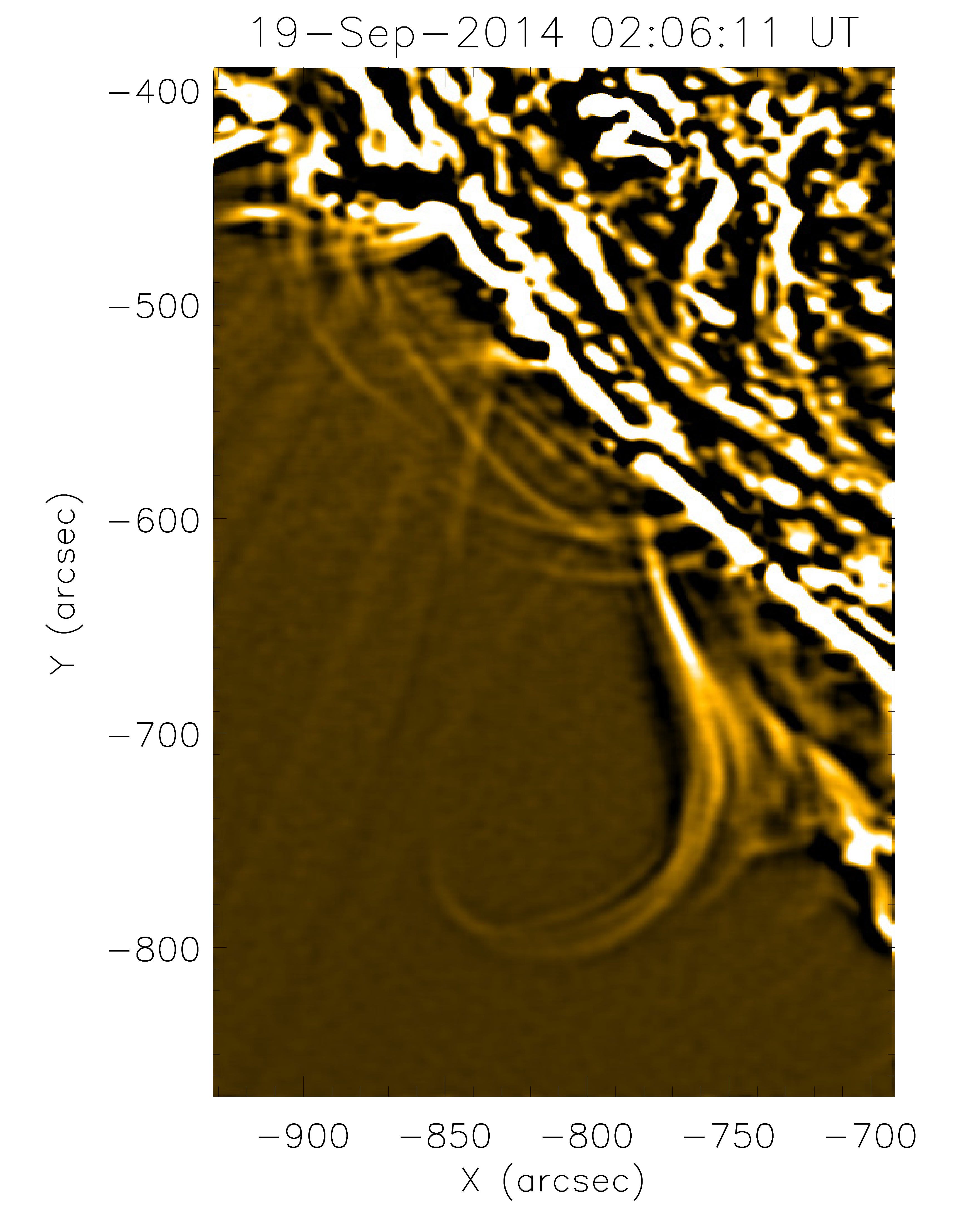}
               \label{clearloop}
               \hspace*{-0.03\textwidth}
               \includegraphics[width=0.515\textwidth,height=10cm,clip=]{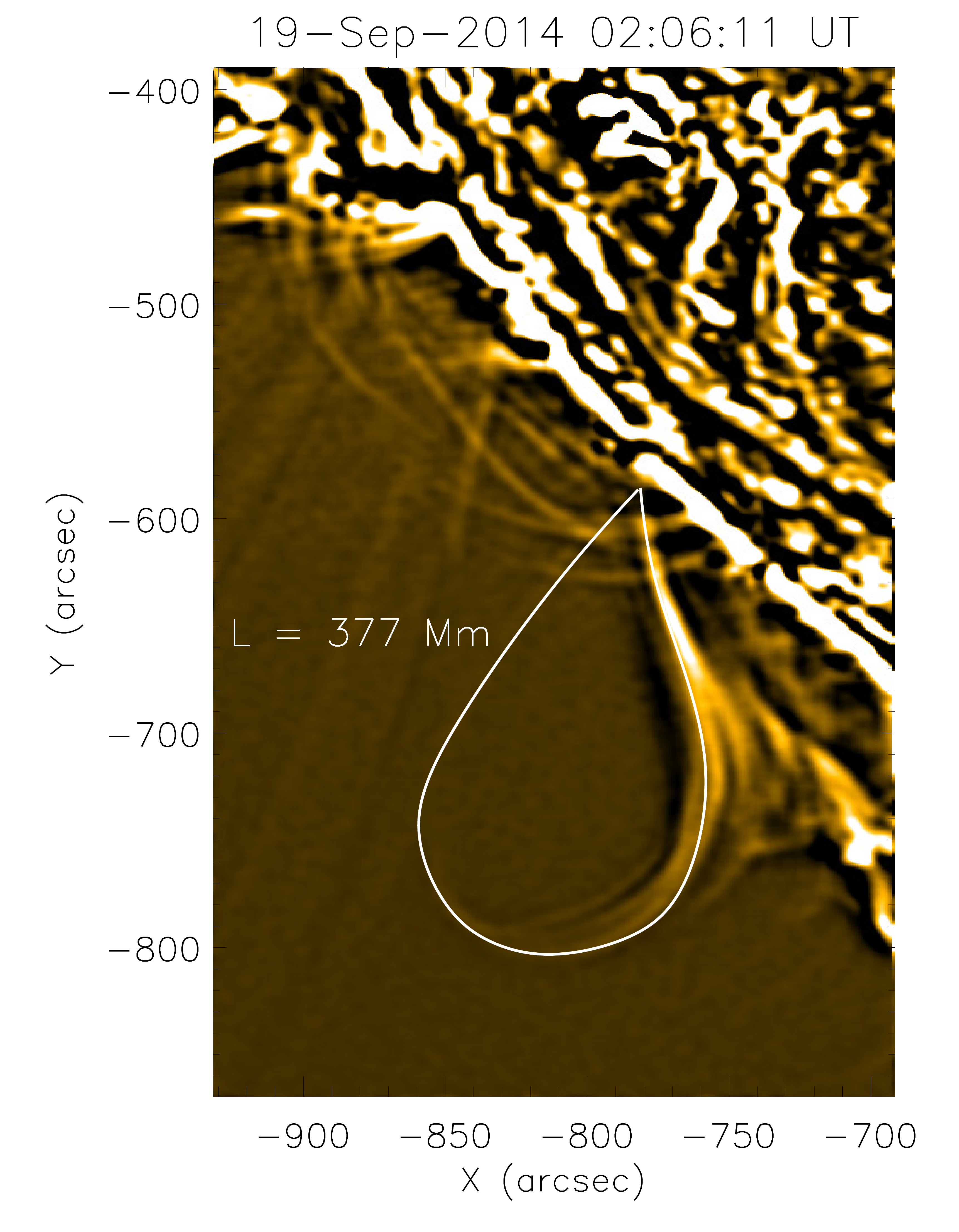}
               \label{clearlooplength}
              }             
\caption{The left panel shows the coronal loop using feature enhancement by the Laplacian operator. The right panel shows the tracing of the loop, using cubic spline fitting.}
\end{figure}
%%%%%%%%%%%%%%%%%%%%%%%%%%%%%%%%%%%%%%%%%%%%%
%%%%%%%%%%%%%%%%%%%%%%%%%%%%%%%%%%%%%%%%%%%%%

In {Figure}~\ref{r2}, it is worth noting that only half of the coronal loop is clearly visible. To enhance the contrast, we smooth the images (to remove noise) and convolve them with the Laplacian operator to enhance the regions of the sharp change of the brightness. The processed image is shown in the left panel of {Figure}~\ref{clearloop}. In order to estimate the length of the loop, we use the image obtained after the application of Laplacian operator and choose ten points along the length of the loop. Then we interpolate a curve passing through these points using cubic spline fitting as shown in the right panel of {Figure}~\ref{clearlooplength}. This procedure is repeated several times to calculate the mean projected length $L$ and the standard deviation of the coronal loop; which is estimated to be $\sim$~377~Mm and 7~Mm, respectively. The length of the coronal loop estimated using this method can be taken as a lower limit of the loop length. Furthermore, we estimate the length of the loop assuming it to be a {semicircle}. We calculate the radius by estimating the distance between the solar limb and the coronal loop top. Since there is uncertainty in the estimation of the coronal loop top (because the coronal loop is quite thick), we take the thickness of the coronal loop as the error in the measurement of the radius. Using this, we estimate the length of the coronal loop to be 539 $\pm$ 30 Mm. This can be taken as the upper limit of the loop length. It is worth noting that the calculated radius is projected in the plane of sky. Thus the estimated length of the coronal loop is still an underestimate of its true length.\\

\subsection{Transverse oscillations in coronal loop}
Seven artificial slices are placed perpendicular to the coronal loop to detect transverse oscillations. Corresponding to each slice, we generate a time-distance map which is referred to as {x--t} map, henceforth, throughout the manuscript.  The {x-axis represents time} and the y-axis represents the distance along the slice.

%%%%%%%%%%%%%%%%%%%%%%%%%%%%%%%%%%%%%%%%%%%%%
%%%%%%%%%%%%%%%%%%%%%%%%%%%%%%%%%%%%%%%%%%%%%
\begin{figure}[h!!!!!!!!!!!!!!!!!!!!!!!!!!] 
  %\begin{left}
 %   \includegraphics[width=.969\textwidth,height=10cm]{rr4_2016_withoutslit8.eps}
    \includegraphics[scale=0.08]{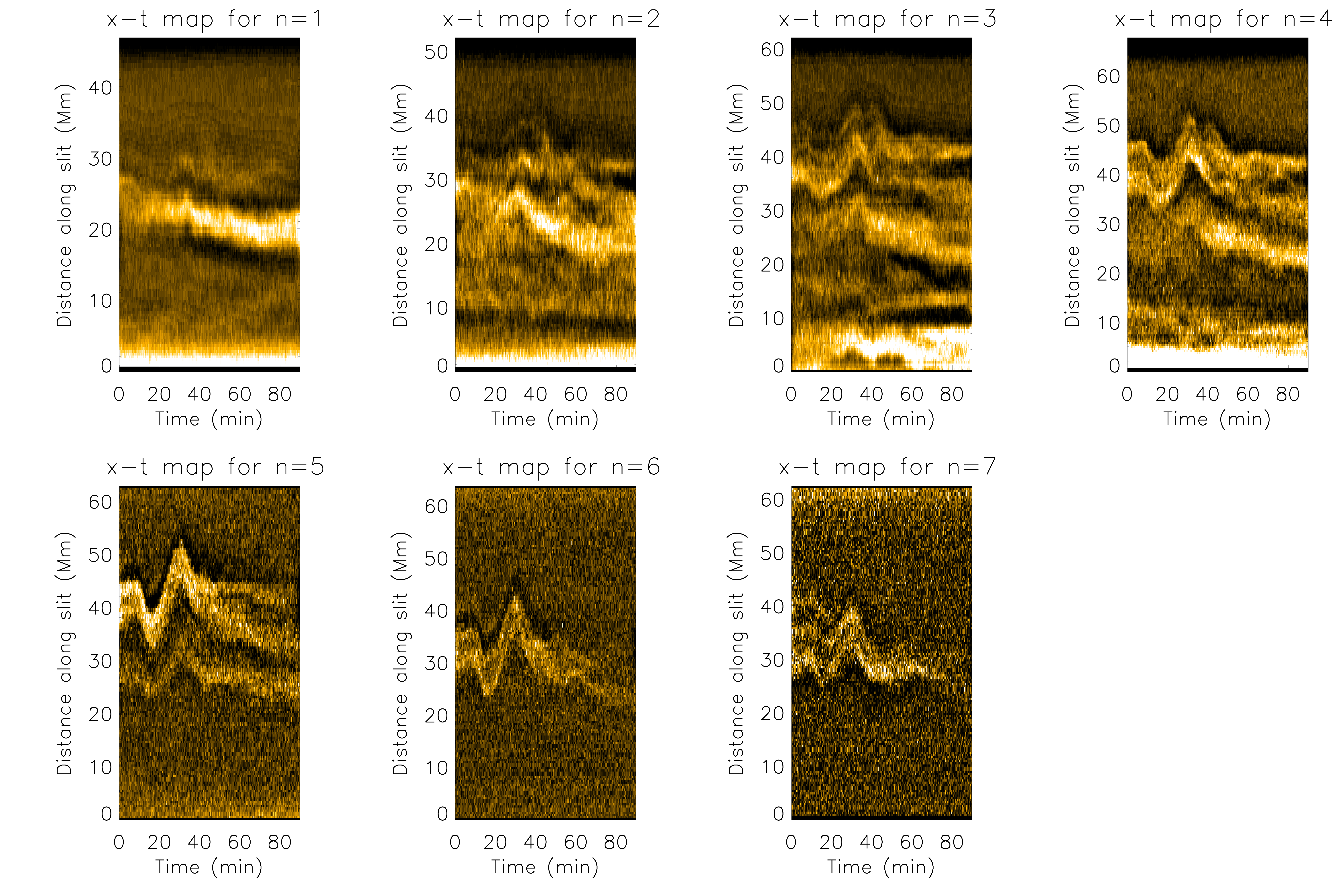}
    \caption{Sharpened {x--t} maps corresponding to the seven slices placed on the loop (see {Figure}~\ref{r2}). {x--t} maps were sharpened using unsharp mask. The {x--t} graph shows the evidence of transverse oscillations in the coronal loop. We also note that the coronal loop is not monolithic but a bundle of fine strands which are oscillating coherently.}
    \label{r4}
 % \end{left}
\end{figure}
%%%%%%%%%%%%%%%%%%%%%%%%%%%%%%%%%%%%%%%%%%%%%
%%%%%%%%%%%%%%%%%%%%%%%%%%%%%%%%%%%%%%%%%%%%%

Time-distance maps are sharpened using the unsharp mask technique. The sharpened maps are shown in {Figure}~\ref{r4}. It is quiet evident from the {x--t} maps that the coronal loop undergoes transverse oscillations.  It is also worth noting that the loop under study is not monolithic but consists of many fine strands, which are oscillating coherently.

\begin{figure}[h!!!!!!!!!!!!!!!!!!!!!!!!]
\centerline{\hspace*{0.015\textwidth}
               \includegraphics[width=0.515\textwidth,trim=0cm 22cm 0cm 0cm, clip=true]{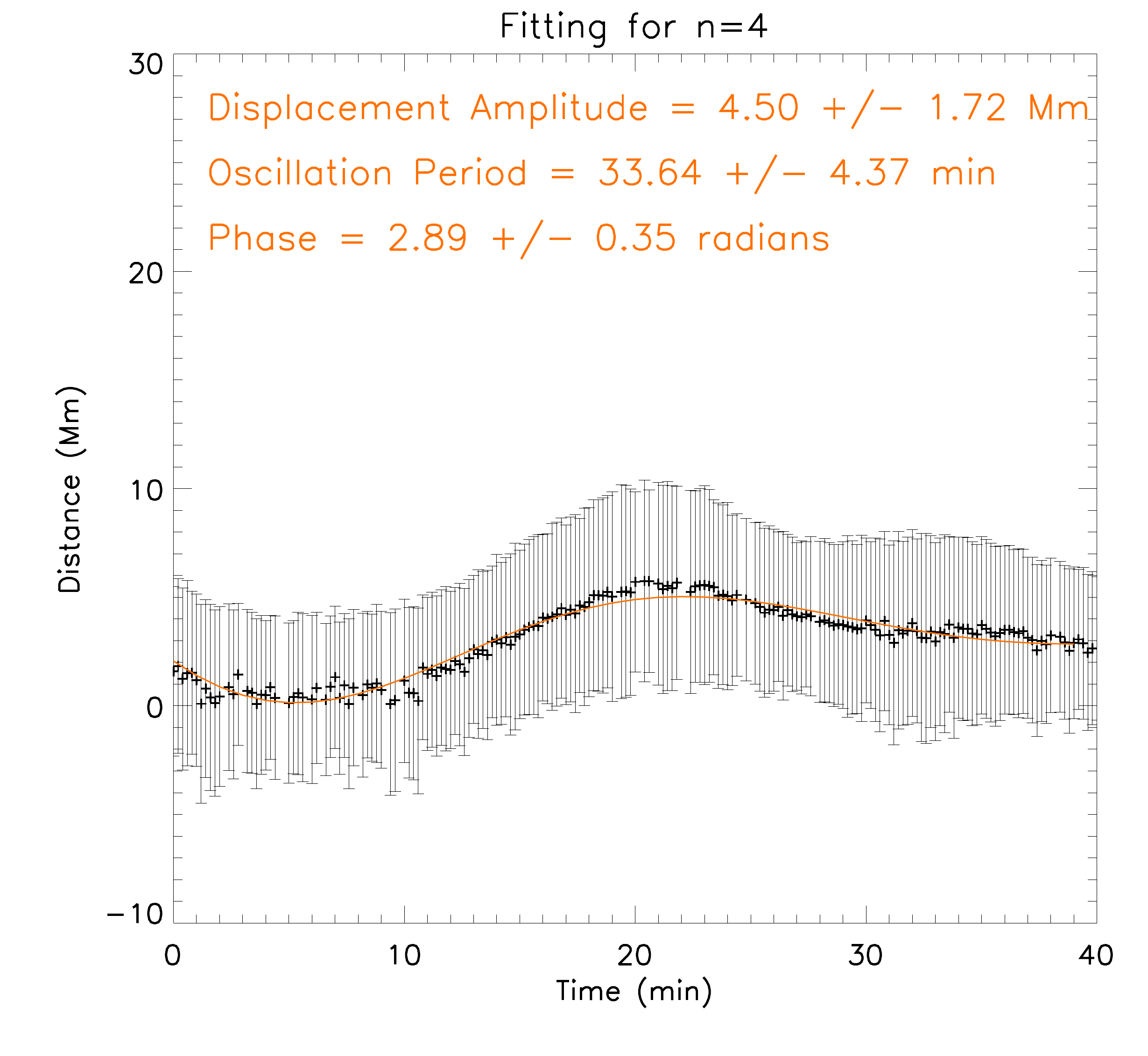}
               \hspace*{-0.03\textwidth}
               \includegraphics[width=0.515\textwidth,trim=0cm 22cm 0cm 0cm, clip=true]{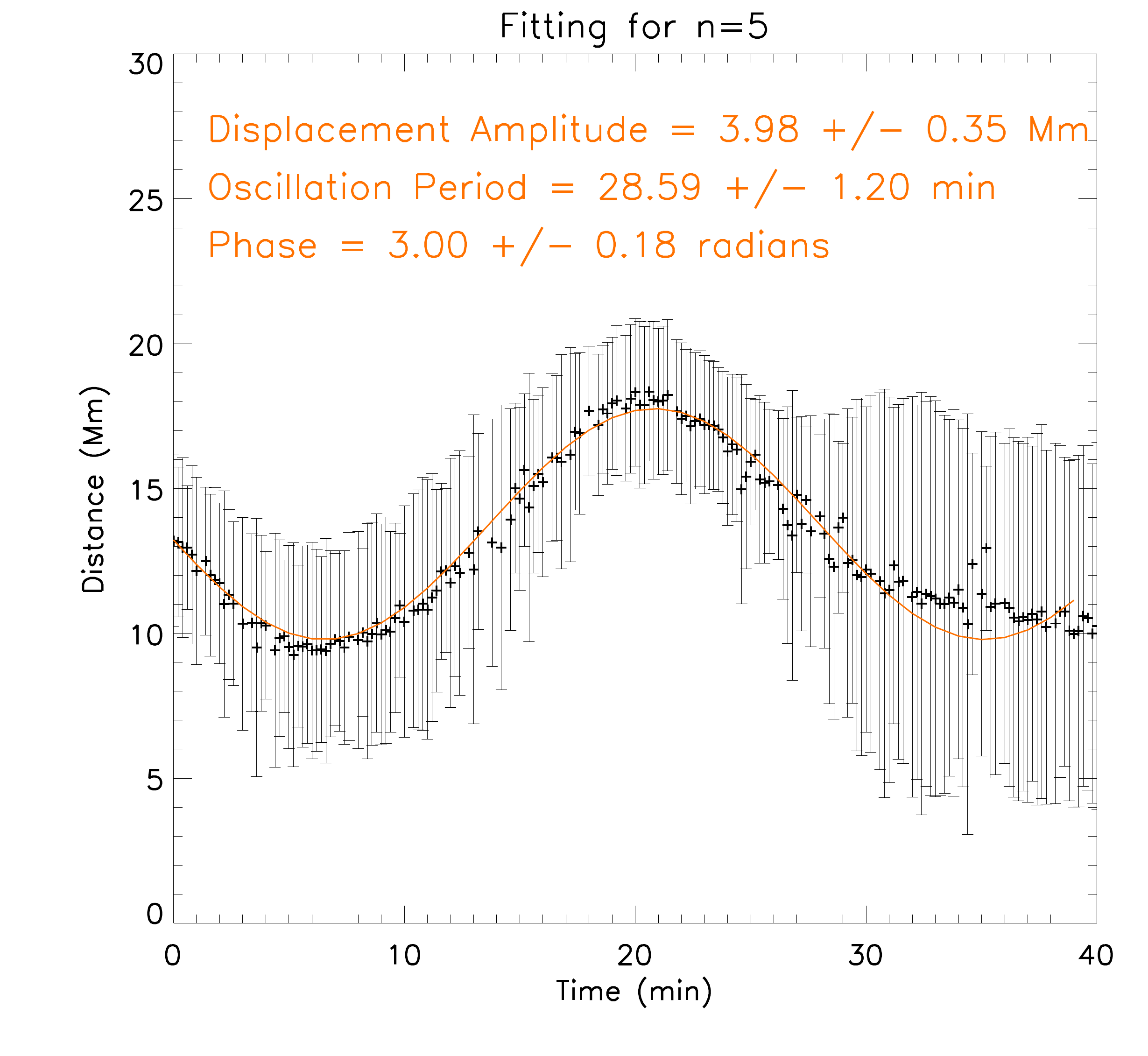}
            }
            
\centerline{\hspace*{0.015\textwidth}
            \includegraphics[width=0.515\textwidth,trim=0cm 22cm 0cm 0cm, clip=true]{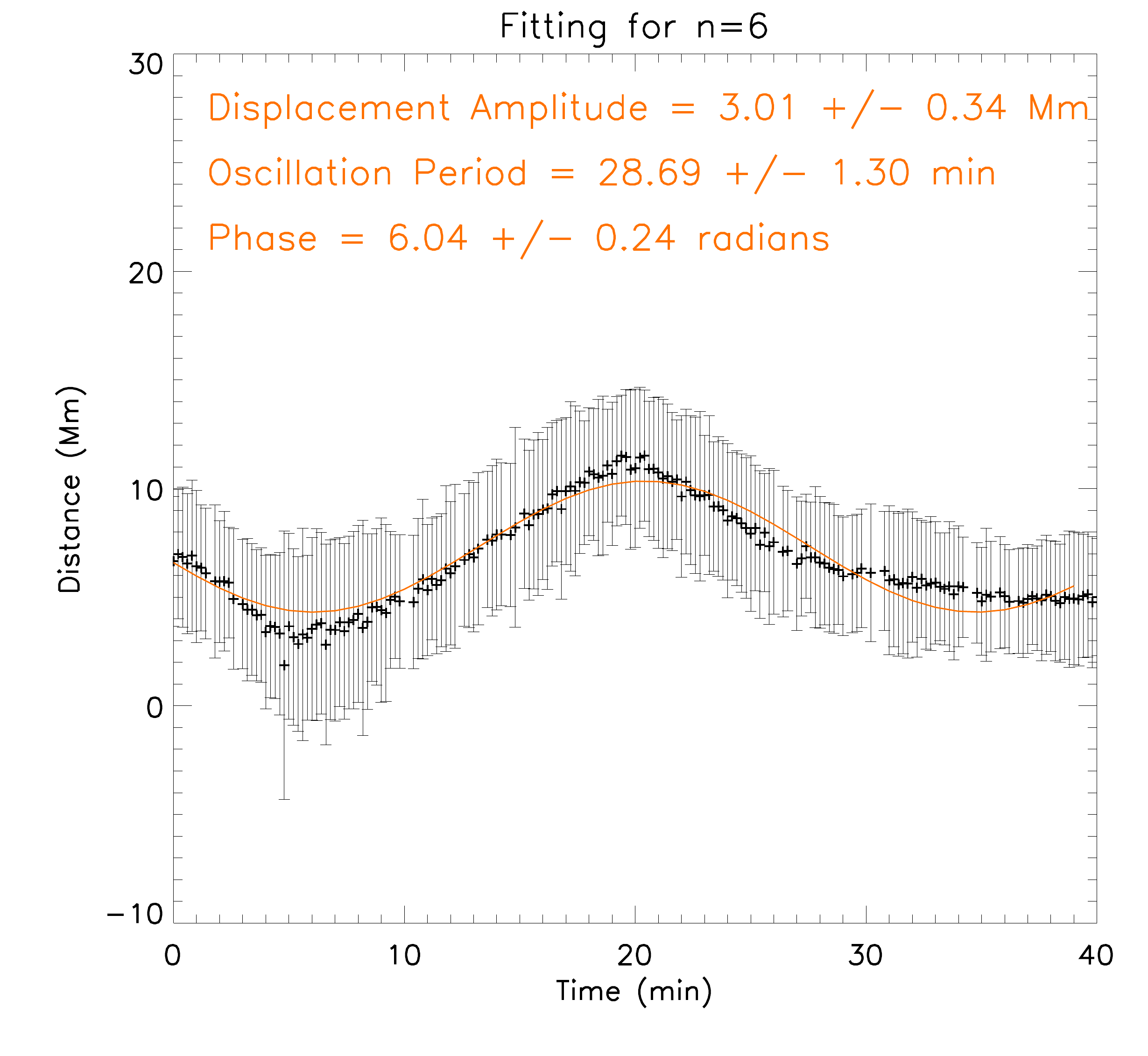}
            \hspace*{-0.03\textwidth}
            \includegraphics[width=0.515\textwidth,trim=0cm 22cm 0cm 0cm, clip=true]{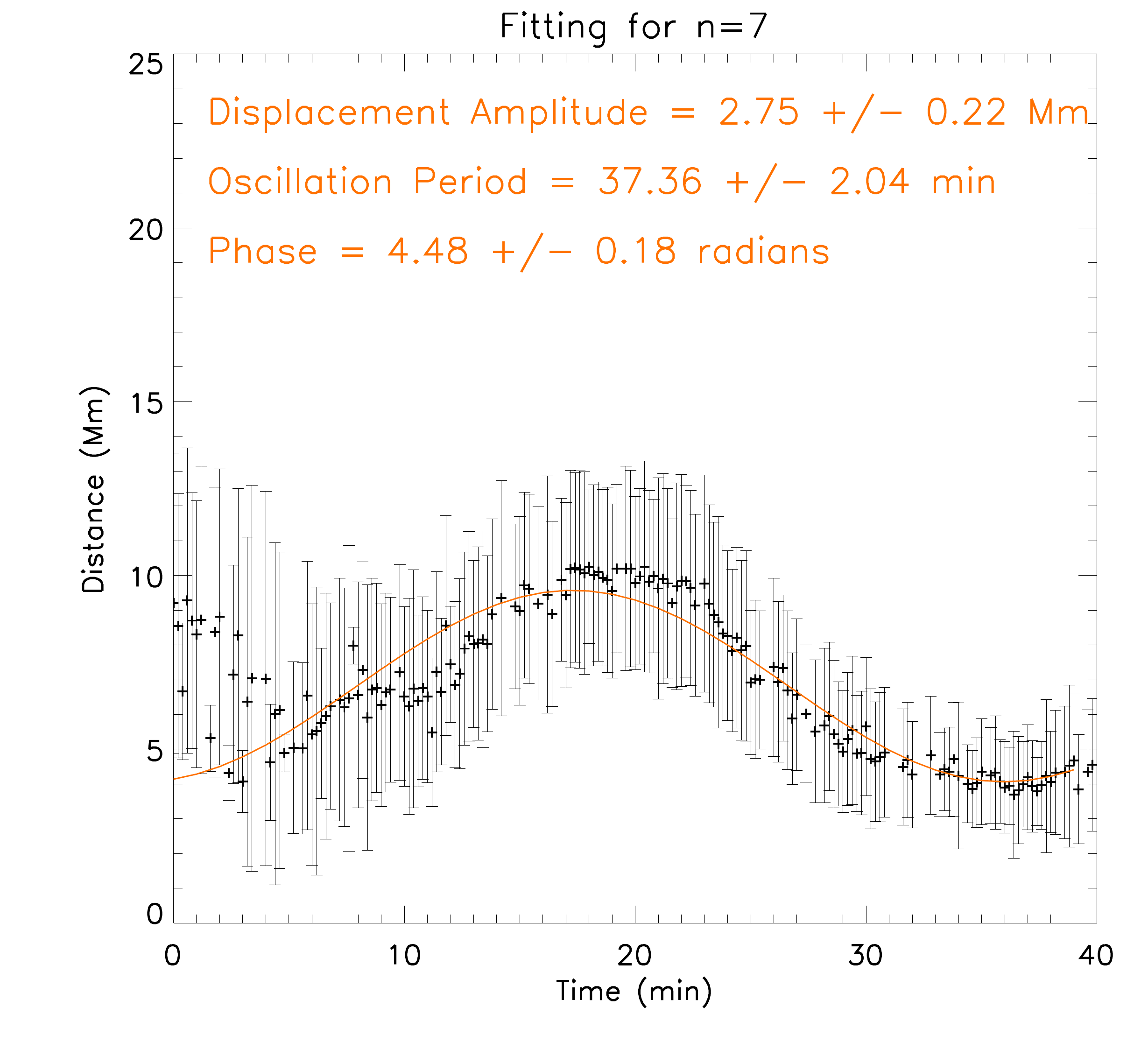}
            }
\caption{Best fit sinusoidal curves with fitting parameters for four slices denoted by n~=~4, 5, 6, 7.}
\label{fitting}
\end{figure}

For the calculation of the dynamic parameters for the coronal loop, first a Gaussian is fitted along each column of a given {x--t} map, and the mean values and one-sigma errors are estimated. We apply this procedure on the {x--t} maps for slice, n~=~4,~5,~6 and~7 only because slices, n~=~1,~2 and ~3 are near to the footpoint of the coronal loop, thus the displacement of the coronal loop is very small. Moreover, near the footpoint of the coronal loop under study, there are many other fine loops along the line of sight that make the detection of transverse oscillation even more difficult. From the {x--t} maps, we notice that the oscillations have a very poor quality factor (only one complete cycle is observed with {no clear gradual} decrease of the amplitude), and therefore we fit the {x--t} maps with an undamped sinusoidal function having the {expression:}
\begin{equation}
A(t)=C+A_0\sin(\omega t+\phi )~,
\label{dse}
\end{equation}
{where} C~=~equilibrium position of the loop, A$_0$~=~displacement amplitude, $\omega$~=~{oscillation} frequency, $\phi$~=~{phase.}\\
We use the MPFIT function \citep{mark2009} in the Interactive Data Language (IDL) to obtain the best fit values of the undamped sinusoidal equation parameters. The best fit sinusoidal curve with best fit parameters are shown in {Figure}~\ref{fitting}.

%%%%%%%%%%%%%%%%%%%%%%%%%%%%%%%%%%%%%%%%%%%%%
%%%%%%%%%%%%%%%%%%%%%%%%%%%%%%%%%%%%%%%%%%%%%
     
     %%%%%%%%%%%%%%%%%%%%%%%%%%%%%%%%%%%%%%%%%%%%%
%%%%%%%%%%%%%%%%%%%%%%%%%%%%%%%%%%%%%%%%%%%%%

     From the fitted curves in {Figure}~\ref{fitting}, we estimate the average oscillation period ($P=\frac{2\pi}{\omega}$) to be $\sim$ 32 $\pm$ 5~min. We note that oscillations at different positions along the coronal loop are in phase. This mode of oscillation is called the \textit{global kink mode}. We also notice that the one foot point of the loop {appears to be anchored at} the limb. Using the Laplacian filtered image ({see Figure}~\ref{clearlooplength}) we find that the other foot point may be anchored {behind the limb}, thus we infer this mode of oscillation as the \textit{fundamental standing mode}.
     
\section{Observations using STEREO}\label{stereo}
\begin{figure}[h!] 
  %\begin{left}
 %   \includegraphics[width=.969\textwidth,height=10cm]{rr4_2016_withoutslit8.eps}
    \includegraphics[scale=0.6]{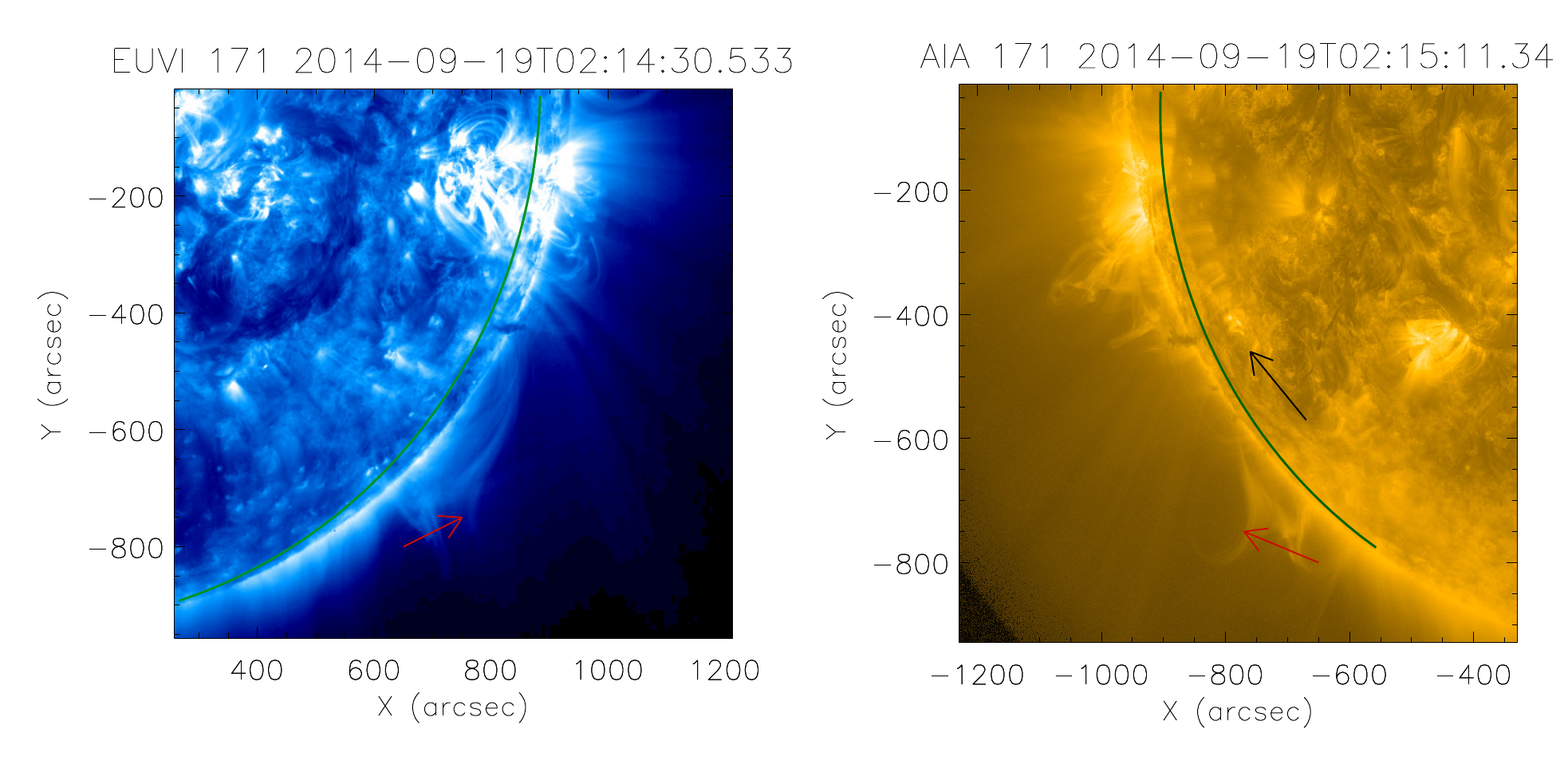}
    \caption{Left: STEREO/EUVI 171~\AA~image  with {the solar limb} as seen from AIA 171~\AA~over plotted in green. The possible position of the loop corresponding to loop seen in AIA 171~\AA~is marked with red arrow. Right: SDO/AIA 171~\AA~image with {the limb} as seen from EUVI 171~\AA~over plotted in green. {The loop} under study is marked with red arrow. Black arrow represents the location of the source region (foot point) of {the jet}. Since the source region is {behind} the limb as seen from EUVI thus it is not seen in EUVI 171~\AA~image.}
    \label{limb171}
 % \end{left}
\end{figure}

\begin{figure}[h!] 
  %\begin{left}
 %   \includegraphics[width=.969\textwidth,height=10cm]{rr4_2016_withoutslit8.eps}
    \includegraphics[scale=0.6]{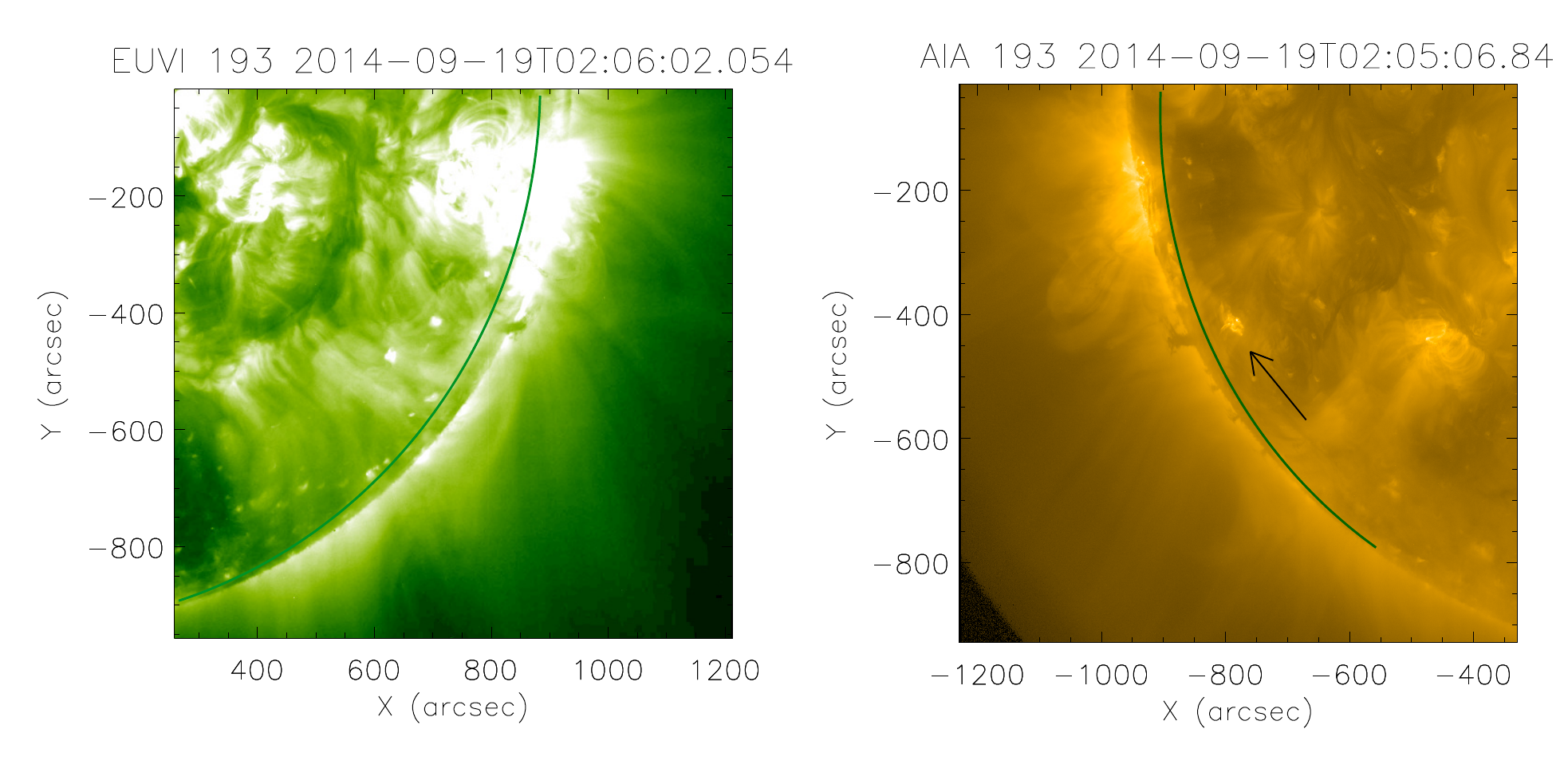}
    \caption{Left: STEREO/EUVI {195}~\AA~image  with {the solar limb} as seen from AIA 193~\AA~over plotted in green. Right: SDO/AIA 193~\AA~image with {the limb} as seen from EUVI 195~\AA~over plotted in green. Black arrow represents the location of the source region (foot point) of {the jet}. Since the source region is {behind} the limb as seen from EUVI thus it is not seen in EUVI 195~\AA~image. The animations are available online as movie~3 and movie~4.}
    \label{limb193}
 % \end{left}
\end{figure}

\begin{figure}[h!] 
  %\begin{left}
 %   \includegraphics[width=.969\textwidth,height=10cm]{rr4_2016_withoutslit8.eps}
    \includegraphics[scale=0.6]{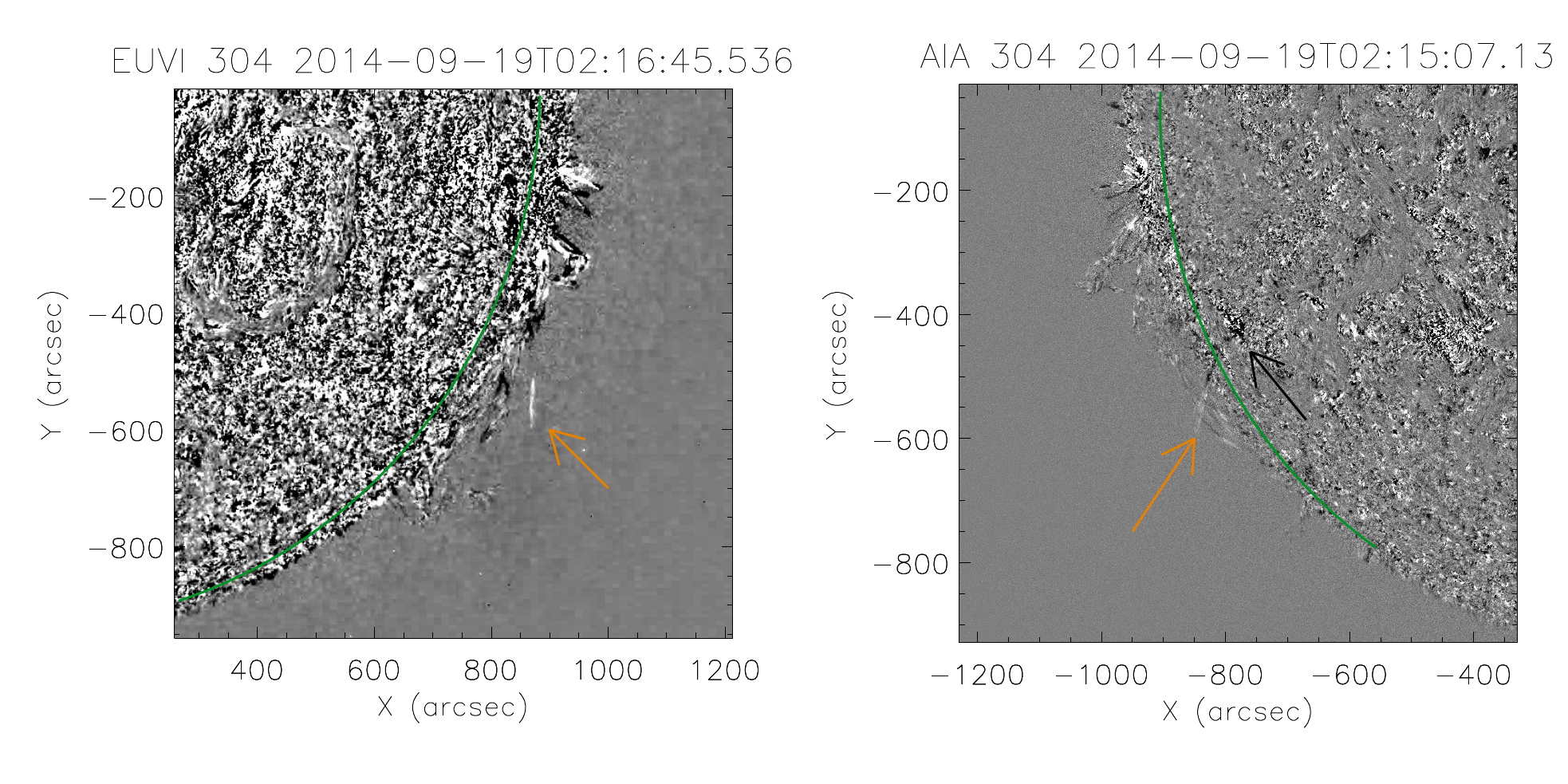}
    \caption{Left: STEREO/EUVI 304~\AA~difference image  with {the solar limb} as seen from AIA 193~\AA~over plotted in green. Right: SDO/AIA 304~\AA~ image with {the limb} as seen from EUVI 195~\AA~over plotted in green. Black arrow represents the location of the source region (foot point) of jet. Arrow in orange represents the jet in EUVI 304~\AA~and AIA 304~\AA. The animations are available online as movie~5 and movie~6. }
    \label{diff_304}
 % \end{left}
\end{figure}

In this section, we use STEREO/{EUVI-B 171, 195, and 304~\AA~}images to get {an} additional perspective of the coronal loop and the jet. We explore if it is possible to get a stereoscopic view of the loop and the jet. In {Figure}~\ref{limb171}, the limb of Sun as seen from AIA is overplotted {on an} EUVI 171~\AA~image and {the solar limb} as seen from EUVI is overplotted {on a} AIA 171~\AA~{image}. We find that the {jet source} region is not observed {in the} EUVI 171~\AA~image because it is {behind} the limb. Since the cadence of EUVI 171~\AA~ is one hour thus we {could not} see the trajectory of the jet in EUVI 171~\AA. In addition to this only few parts of the coronal loop under study are seen in EUVI 171~\AA. The possible position of the loop under study is marked with a red arrow in EUVI~171~\AA. The corresponding position in AIA~171~\AA~ is also marked with a red arrow. We could not associate the {loop features} as seen from EUVI 171~\AA~ with the corresponding {loop features} as seen from AIA 171~\AA. For example, the loop we see in EUVI 171~\AA~(marked with red arrow in EUVI~171~\AA), could be the other foot point of the coronal loop which is not clearly seen in AIA 171~\AA~image (marked with red arrow in AIA 171~\AA). Thus it is extremely difficult to perform the triangulation in such a scenario. We repeat the same analysis with EUVI 195~\AA~ images (see {Figure}~\ref{limb193}). Since EUVI 195~\AA~has a better cadence of 5 min, we {are able to} see the jet in one frame in EUVI 195. We have made a normal intensity movie and a difference movie (movie~3 and movie~4) to illustrate it. Similarly the jet is also seen in EUVI 304~\AA~difference images ({see Figure}~\ref{diff_304}). We have also made a normal intensity and a running difference movie (movie~5 and movie~6) using EUVI 304~\AA~and AIA 304~\AA.

\subsection{EUVI 195~\AA}
\label{euv195}
\begin{figure}[h!] 
  %\begin{left}
 %   \includegraphics[width=.969\textwidth,height=10cm]{rr4_2016_withoutslit8.eps}
    \includegraphics[scale=0.6]{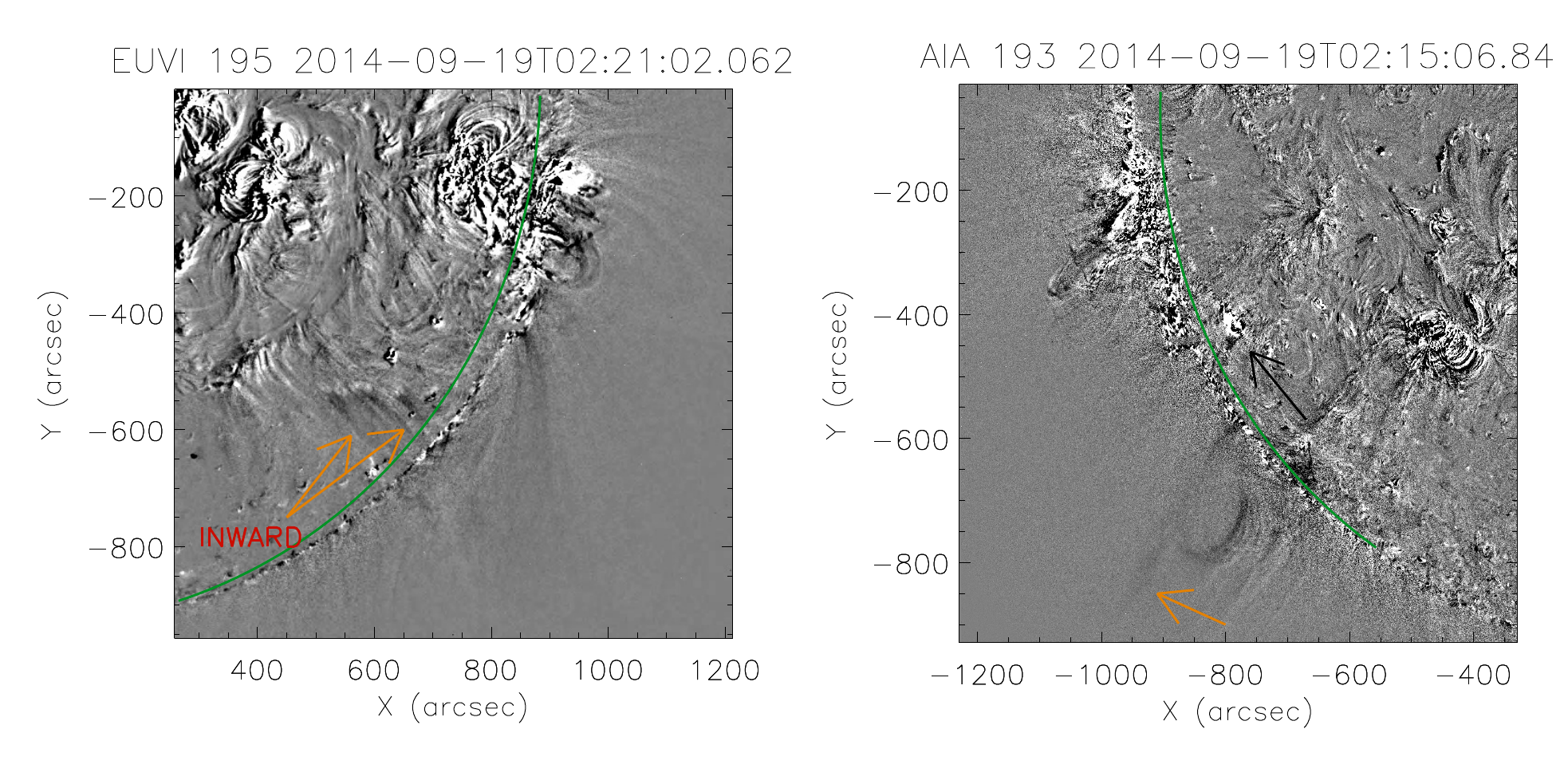}
    \caption{Left: STEREO/EUVI 195~\AA~ difference image  with {the solar limb} as seen from AIA 193~\AA~over plotted in green. Right: SDO/AIA 193~\AA~ image with {the limb} as seen from EUVI 195~\AA~over plotted in green. Black arrow represents the location of the source region (foot point) of {the jet}. {Arrows} in orange {represent} the dark feature in EUVI 195~\AA~and AIA 193~\AA. }
    \label{diff_193}
 % \end{left}
\end{figure}

{In the} difference movie of EUVI 195~\AA~and AIA~193~\AA~(movie~4), a disturbance is seen to be moving inwards towards the solar disk in EUVI 195~\AA. On close inspection of EUVI 195~\AA~images (see movie~3), we note that the plasma disappears at the location where we see a dark feature in difference {images}. It starts near the limb and propagates inward towards the disk (marked in {Figure}~\ref{diff_193} (left panel)). This disappearance of plasma {manifests} as a dark feature in EUVI~195~\AA~difference images ({see Figure}~\ref{diff_193}). {This} is unlikely to be caused by an eruption because otherwise it would have been propagating outwards. There could be few possible scenarios, one that the disappearance might have been due to the heating of the plasma from a nearby flare or some other event (jet in this case) which transfers the energy to the surrounding medium. Secondly,  there can also be a change in the magnetic field topology due to magnetic reconnection which may cause plasma to escape through open field lines and therefore {appearing} as a dark feature in EUV 195~\AA. Furthermore, it is also possible that the inward feature that is seen in EUVI-B 195~\AA~is the jet material falling  down to the surface following the local magnetic field lines. We notice from movie~4 that the dark feature first appears in AIA 193~\AA. It interacts with the coronal loop and displaces it. Subsequently it appears in EUVI 195~\AA~and propagates inwards. Therefore  disturbance seen in EUVI~195~\AA~must be associated {with the} jet as seen in AIA.    

 \section{Analysis of the jet}\label{sec4}
High resolution images obtained from AIA/SDO enable us to resolve the jet and characterise its properties. Often jets are observed near coronal holes \citep{chandra2014}. Appearance and dynamics of jets have been well studied due to their contribution to coronal heating and solar wind acceleration \citep{chandra2014,mueller2008,savcheva2007,shibata1995}.

\begin{figure}[h!] 
  %\begin{left}
 %   \includegraphics[width=.969\textwidth,height=10cm]{rr4_2016_withoutslit8.eps}
    \includegraphics[scale=0.6]{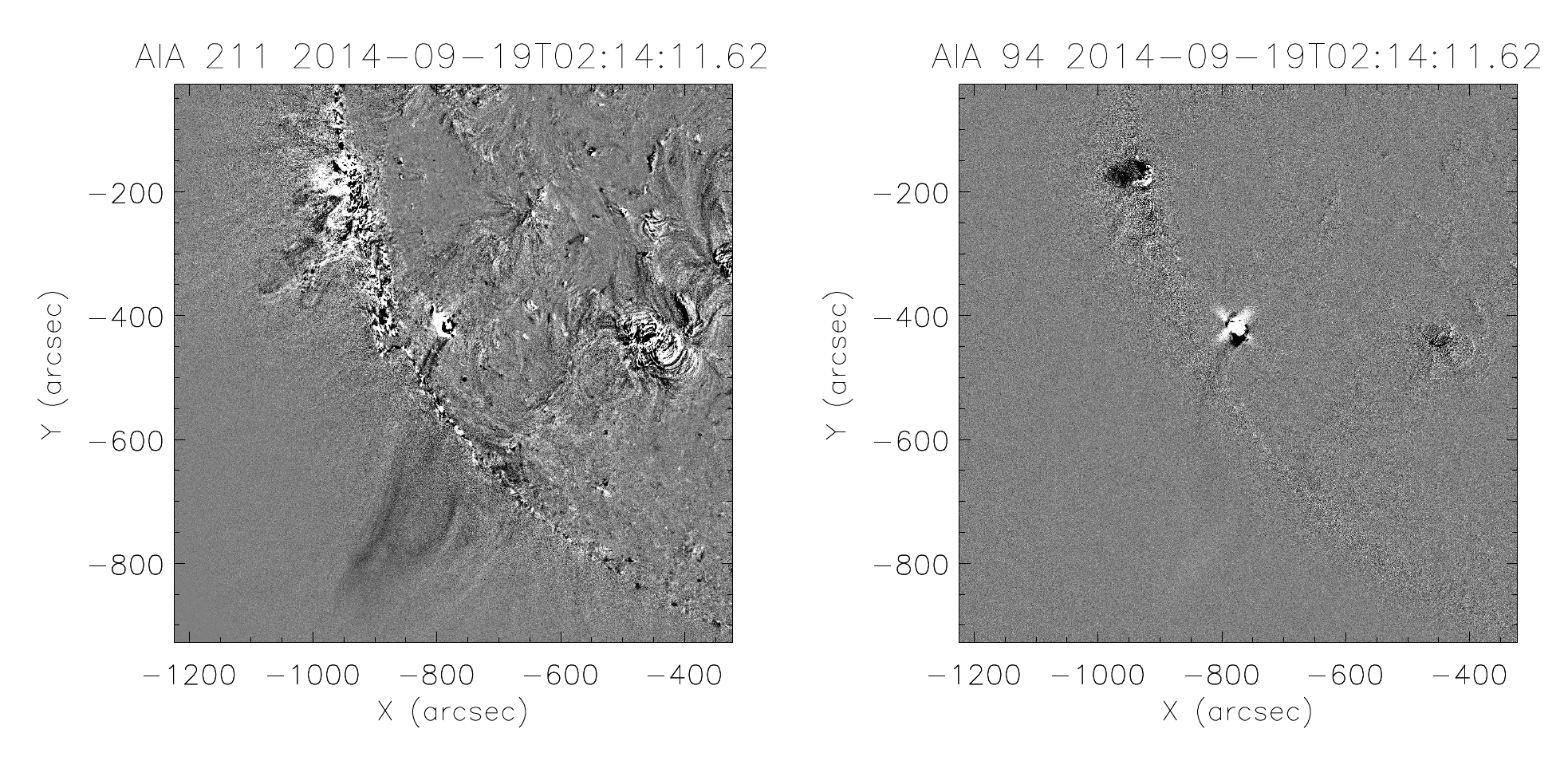}
    \caption{Left: SDO/AIA 211~\AA~ difference image. A dark feature is seen near the loop. Right: SDO/AIA 94~\AA~ image. A faint bright emission is seen near the loop. The animations are available online as movie~7 and movie~8. }
    \label{diff_211_94}
 % \end{left}
\end{figure}

The jet is seen in AIA~304~\AA, 171~\AA, 193~\AA, 211~\AA~ and to some extent in 94~\AA. In AIA 304~\AA~and 171~\AA~ the jet is seen as a bright collimated plasma propagating outward while in hotter channels like AIA 211~\AA ~and to some extent in 193~\AA, we notice bright collimated plasma initially, which becomes fainter at later times, followed by a dark emission which is most likely due to the heating of the plasma. In 94~\AA~we see a faint emission which suggests that the jet is multithermal in nature having both cool and hot components (see movie~7 and movie~8). As explained above in AIA 211~\AA~ and 193~\AA, a dark feature {seen} in difference images which is the manifestation of the dark emission seen in normal intensity images. At the same location in AIA 94~\AA~a faint emission is seen, which is the signature of heating (see {Figure}~\ref{diff_211_94}). The animations of difference images of AIA 211~\AA~and 94~\AA~are available online as movie~7 and movie~8, respectively. It is worth {noting} at this point that the dark feature seen in AIA 211~\AA, either could be the hot component of {the jet} or it could be the signature of heating caused by {the jet}. However, due to low signal to noise we {could not} find bright emission in 94~\AA~but certainly a faint emission is seen in the difference images. The mechanism of heating is beyond the scope of present study.

\begin{figure}[h!] 
  %\begin{left}
 %   \includegraphics[width=.969\textwidth,height=10cm]{rr4_2016_withoutslit8.eps}
    \includegraphics[scale=0.6]{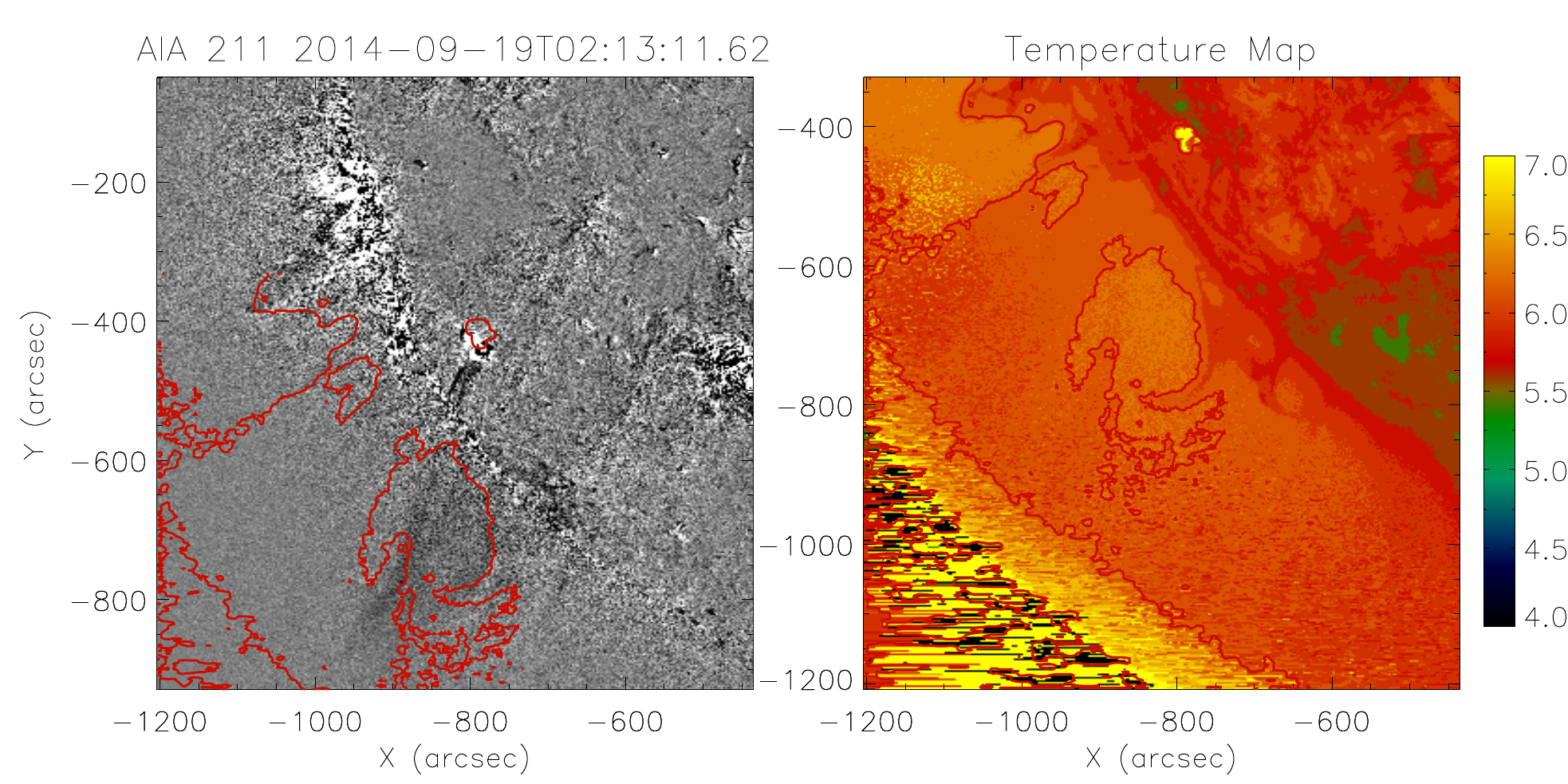}
    \caption{Left: SDO/AIA 211~\AA~ difference image with contours of log(T)=6.3 ($\sim 2$ MK) overplotted. Right: {Temperature map (note the difference in field of view)}. A colorbar {on the} right {represent} log(T).}
    \label{diff_211t}
 % \end{left}
\end{figure}

We found that the jet is associated with a B9 class flare as recorded by GOES-15, {whose} peak is recorded at 02:13 UT. It may also be possible that the inward moving dark features observed with STEREO/EUVI-B 195~\AA~could be the faint EUV wavefront triggered by the jet \citep{liu2014} or the result of jet material falling {back to} the solar surface \citep{culhane2007}.
%Moreover we also note that the dark emission (which appears as dark feature in difference images) is propagating faster and hence ahead of bright emission seen in AIA 304~\AA ~(see figure~\ref{diff_193_304} and movie~9).  \\
We also estimate the temperature at the location of dark emission using DEM technique developed by \inlinecite{aschdem}. We find at the location of dark emission seen in AIA 211~\AA~ difference images, the temperature is hot as compared to other regions. {Figure}~\ref{diff_211t} right panel shows the temperature map with contours of log(T)=6.3 overplotted and left panel shows the difference image of AIA~211~\AA~ with contours of log(T)=6.3 overplotted in red color. We note that the contours surround most of the dark emission seen in AIA 211~\AA. We conjecture that the jet interacts with the loops, causing transverse oscillations and with the ambient medium transferring part of its energy and heating local plasma. This heating causes plasma to disappear from AIA 193 and 211~\AA~ passbands thus dark features appear {in the hotter channel}. When this heating {shifts towards the far side,beyond the limb}, we see it as {a} dark feature propagating inwards in EUVI 195~\AA. Moreover, as pointed out in {Section}~\ref{euv195}, the propagating dark feature seen in EUVI 195~\AA~could be the jet plasma {falling back} towards the Sun.

\begin{figure} 
\centerline{\hspace*{0.015\textwidth}
               \includegraphics[height=10cm,width=6.5cm,clip=]{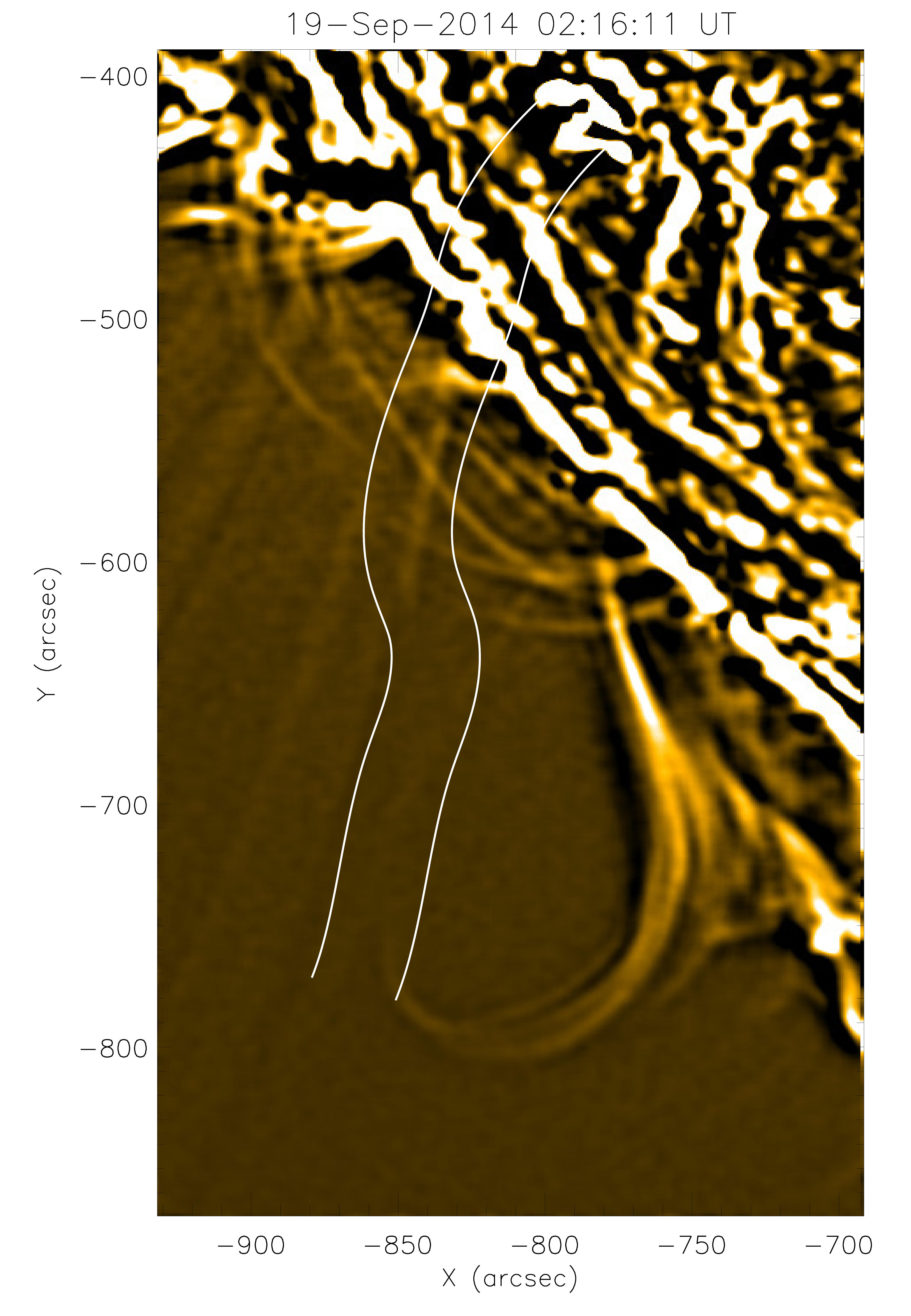}
               \hspace*{-0.03\textwidth}
               \includegraphics[height=8.1cm,width=6.5cm,clip=]{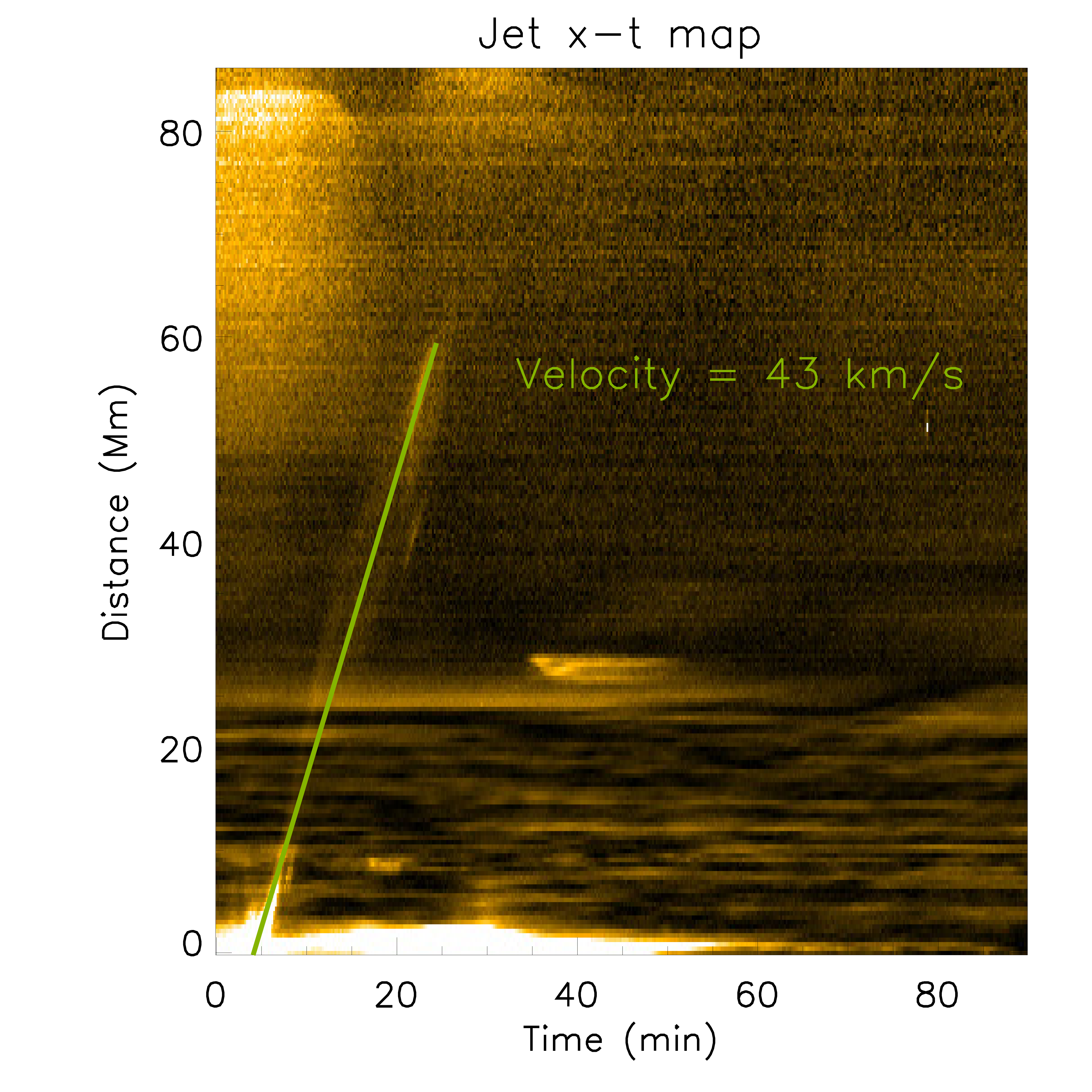}
              }
     
\caption{{The left} panel shows the region of jet propagation is outlined by two parallel curves. Right panel represents the {x--t} map for the region selected (as shown in the left panel). The fitted straight line is shown in green colour. The velocity {of the} jet is found to be $\sim$ $43\pm4$ km s$^{-1}$. An {animation} is available online as movie~2.}
\label{j22}     
\end{figure}

\begin{figure}[h!] 
  %\begin{left}
 %   \includegraphics[width=.969\textwidth,height=10cm]{rr4_2016_withoutslit8.eps}
    \includegraphics[scale=0.1]{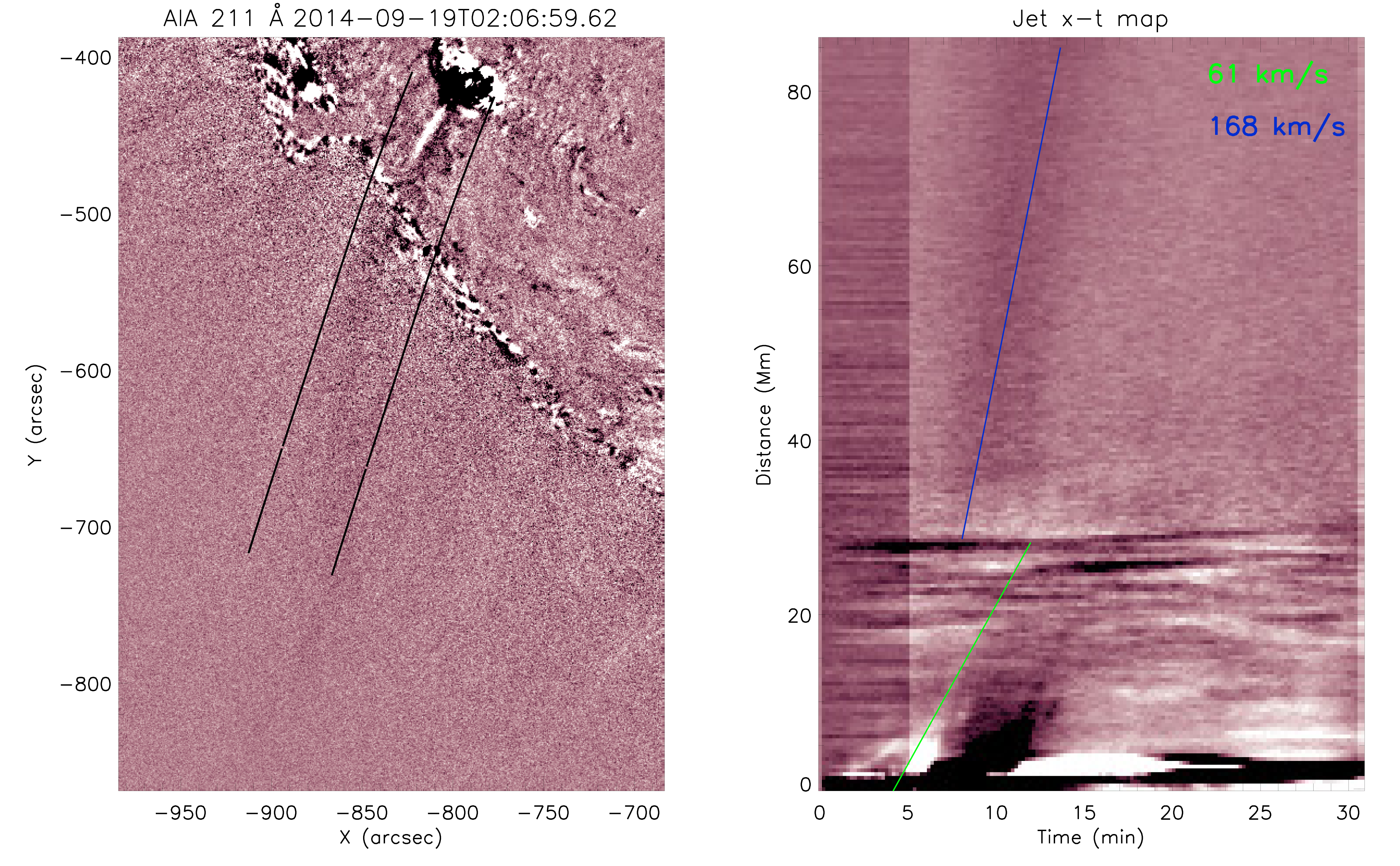}
    \caption{Left: SDO/AIA 211~\AA~ difference image. The jet trajectory is outlined by two parallel lines. Right: {x--t} map for the region selected (as shown in left panel). Two features, bright ridge and dark ridge are fitted with green and blue line respectively.}
   \label{211xt}
 % \end{left}
\end{figure}

We estimate the velocity of {the jet} by creating {an x--t} map. In order to generate the {x--t} map, we isolate the trajectory of the jet. We choose {a few} points along the path through which the jet propagates (starting from the time of jet eruption {until} the jet vanishes). Next, we interpolate a curve between these points using cubic spline fitting. We plot another curve, 50 pixels apart, parallel to it, so that the jet propagates within the region outlined by the two white curves as shown in the left hand panel of {Figure}~\ref{j22}, {in spite} of the transverse displacement of the jet. The trajectory of the jet, marked with white curves, is shown in a movie available online (see movie~2). Finally, we average the intensity between two curves and create the {x--t} map. The {x--t} map is shown in the right panel of {Figure}~\ref{j22}. Inclined ridge in {x--t} map represents the jet propagating outwards along the chosen curved artificial slice. We fit a straight line using eye estimation. The velocity of the jet is estimated by calculating the slope of the fitted line, which is found to be $\sim$~$43\pm4$~km~s$^{-1}$. We follow the same analysis on AIA 211~\AA~difference images and generate {x--t} map as shown in {Figure}~\ref{211xt}. In this case we choose a broad straight slice, to ensure that both bright and dark features remain within the slit {for the} entire duration of the analysis. In {the} right panel of {Figure}~\ref{211xt}, we see bright and dark ridges. The bright ridge corresponds to jet material and is estimated to be propagating with a speed of 63~km~s$^{-1}$. The dark emission is estimated to be propagating with a speed of {168~km~s$^{-1}$}. This could either be the hot component of jet or the heating caused by jet eruption, which is propagating outward. However, due to poor signal to noise in AIA 94~\AA, we could not make {x--t} maps of AIA 94~\AA.

%%%%%%%%%%%%%%%%%%%%%%%%%%%%%%%%%%%%%%%%%%%%%
%%%%%%%%%%%%%%%%%%%%%%%%%%%%%%%%%%%%%%%%%%%%%

%%%%%%%%%%%%%%%%%%%%%%%%%%%%%%%%%%%%%%%%%%%%%
%%%%%%%%%%%%%%%%%%%%%%%%%%%%%%%%%%%%%%%%%%%%%

\begin{figure}
\begin{center}
\centerline{\hspace*{0.015\textwidth}
               \includegraphics[scale=0.09, clip=true]{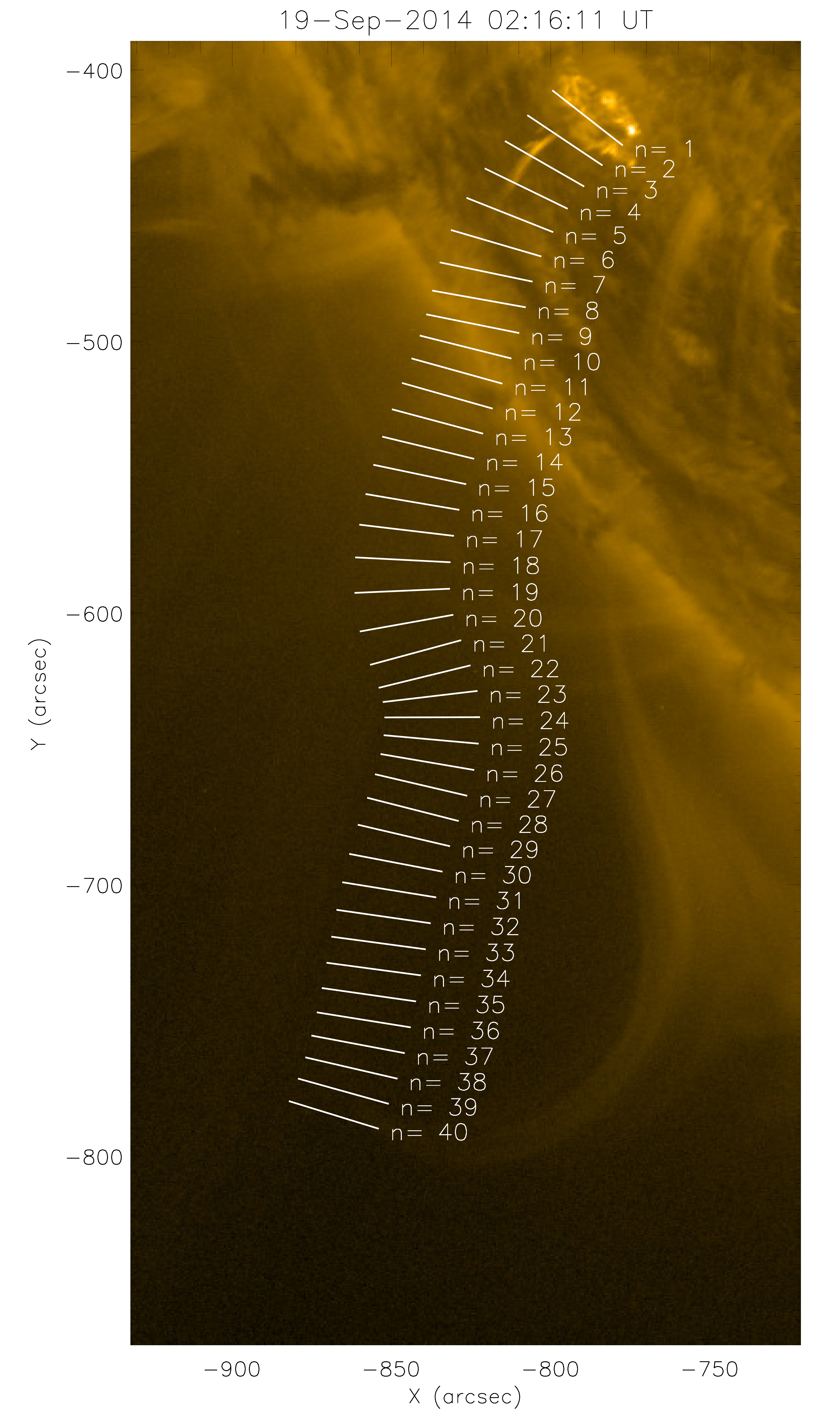}
               \hspace*{-0.03\textwidth}
               \includegraphics[scale=0.15, clip=true]{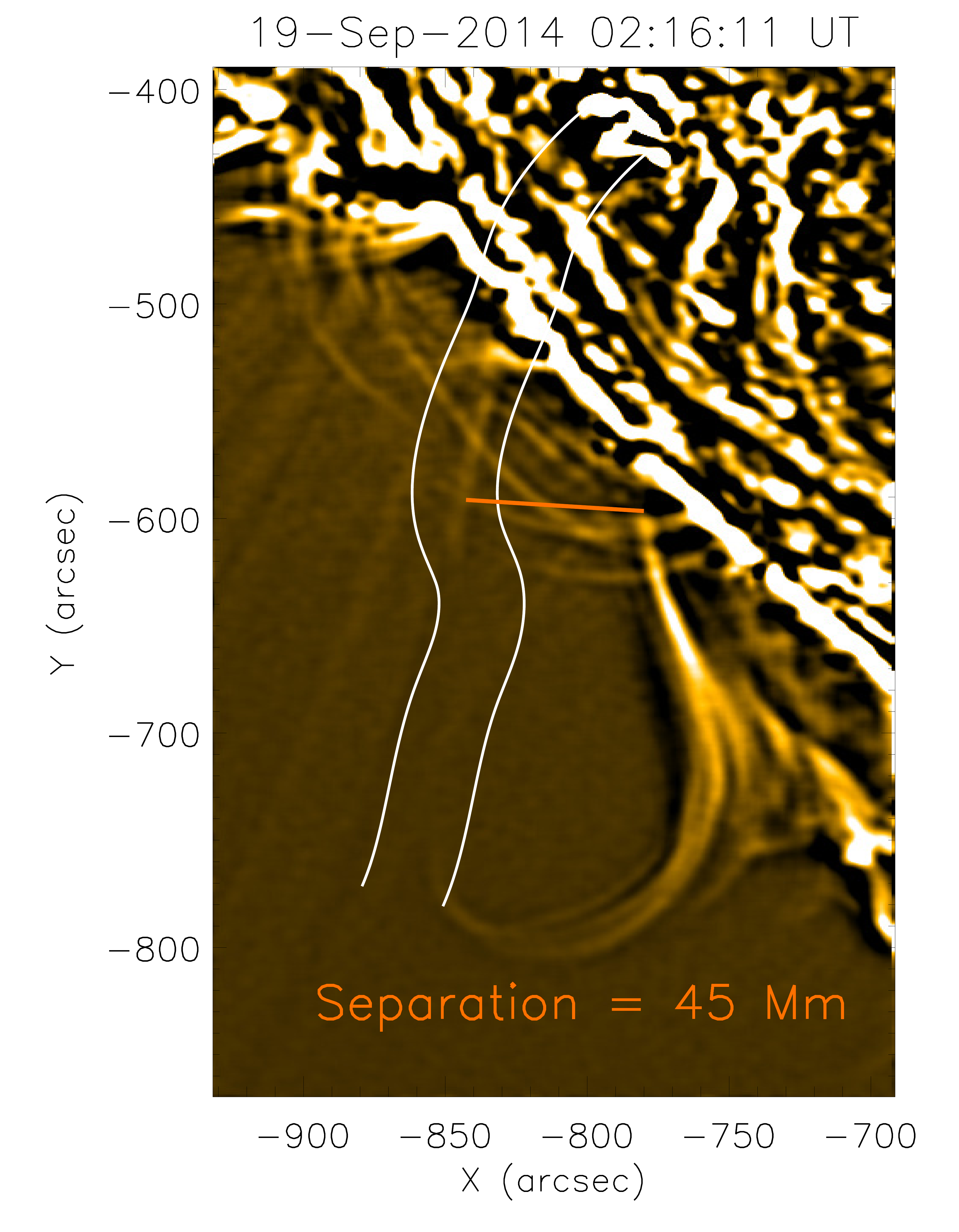}
            }

   \caption{Left: The figure shows 40 transverse slits at 40 equidistant points along the axis of jet propagation. Right: Figure showing the separation between the oscillating jet and coronal loop.}
    \label{jet_s}
      \end{center}
\end{figure}

We also analyse the transverse oscillations in {the} jet as it propagates. Since {the} jet propagates while it oscillates, we placed 40 artificial slices along the axis of jet propagation to capture the transverse oscillations. This is done by joining 40 equidistant points of one of the curves, to its corresponding parallel point on the other curve. Therefore, we get 40 transverse slits at 40 equidistant position along the axis of jet as shown in {Figure}~\ref{jet_s}. 
%This is shown in {Figure}~\ref{jet_s}.

%%%%%%%%%%%%%%%%%%%%%%%%%%%%%%%%%%%%%%%%%%%%%
%%%%%%%%%%%%%%%%%%%%%%%%%%%%%%%%%%%%%%%%%%%%%
%%%%%%%%%%%%%%%%%%%%%%%%%%%%%%%%%%%%%%%%%%%%%
%%%%%%%%%%%%%%%%%%%%%%%%%%%%%%%%%%%%%%%%%%%%%

%{Figure}~\ref{jet_xt} shows 36 {x--t} maps corresponding to the 36 slices placed (refer {Figure}~\ref{jet_s}). The last 4 {x--t}~maps corresponding to the respective 4 slice positions showed no trace of jet and hence were not plotted in the figure.\\
%From the {x--t} maps we can conclude that the motion of the jet is downward first (till it reaches slice 17), then upwards (till it reaches slice 23) and then again downwards (till it reaches slice 27), after which the jet vanishes. Thus the jet undergoes \textit{transverse oscillations} while \textit{propagating}.

%%%%%%%%%%%%%%%%%%%%%%%%%%%%%%%%%%%%%%%%%%%%%
%%%%%%%%%%%%%%%%%%%%%%%%%%%%%%%%%%%%%%%%%%%%%
\begin{figure}
  \begin{center}
    \includegraphics[scale=0.3]{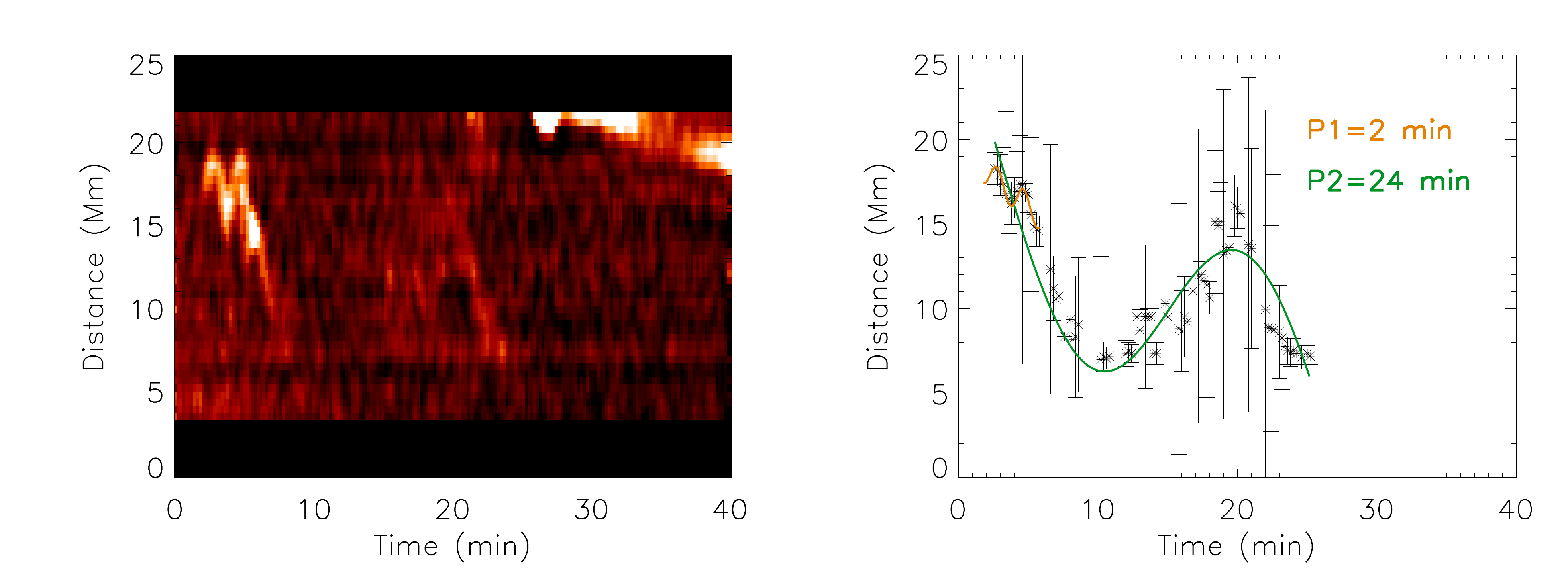}
   \caption{Left: {x--t} map obtained after adding all artificial slices as shown in {Figure}~\ref{jet_s}. Right: Orange and green curves represent the best fit sinusoidal curves. Corresponding periods are mentioned in the figure.}
   \label{jet_xt}
     \end{center}
     \end{figure} 
     
 %%%%%%%%%%%%%%%%%%%%%%%%%%%%%%%%%%%%%%%%%%%%%
%%%%%%%%%%%%%%%%%%%%%%%%%%%%%%%%%%%%%%%%%%%%%

We create {x--t} maps for individual {slices} and add them. The resulting {x--t} map is shown in {Figure}~\ref{jet_xt}. Since the jet gets fainter as it propagates outward, the jet intensity decreases with time in {x--t} maps. We find that both high and low periodicities are co-existent. The high frequency signal however is present only in a region close to the footpoint of the jet. We fit two sinusoidal curves in {Figure}~\ref{jet_xt} as shown in white and green curves and found the periodicity of 2~min and 24~min respectively. It is important to note that the time period of {jet oscillation} ({$\sim24$ min}) is nearly equal to the time period of oscillations of the coronal loop ({$\sim31$ min}). 
     
 %%%%%%%%%%%%%%%%%%%%%%%%%%%%%%%%%%%%%%%%%%%%%%%%
 %%%%%%%%%%%%%%%%%%%%%%%%%%%%%%%%%%%%%%%%%%%%%%%%

\section{MHD Seismology}\label{mhd}
We assume the coronal loop to be a cylindrical magnetic flux tube of uniform magnetic field ${B_0}$, % Let the plasma inside the coronal loop to be of uniform density \textit{$\rho$$_{0}$} and pressure \textit{p$_{0}$}. This configuration of the flux tube is preserved by an external magnetic field \textit{B$_{e}$} having uniform plasma of density \textit{$\rho$$_{e}$} and pressure \textit{p$_{e}$}. {Figure} 13 illustrates this. Let for the external and internal media, the speed of sounds be \textit{C$_{se}$} and \textit{C$_{s0}$}, Alfv\'en speeds \textit{C$_{Ae}$} and \textit{C$_{A0}$}, tube speeds \textit{C$_{Te}$} and \textit{C$_{T0}$}. For both the plasmas to stay in equilibrium the following condition needs to be satisfied,\citep{nv} : 
  %\begin{equation}    
  % p_0+\frac{{B_0}^2}{2\mu _0}=p_e+\frac{{B_e}^2}{2\mu _0}
  %\end{equation}   
where the minor radius \textit{a}, that is the half width of the coronal loop cross-section is $\ll$ $L$ (where $L$ is the total length of the loop). This is called the thin tube or long wavelength approximation. Thus, $k_za<<1$ (where, $k_z=2\pi/\lambda$ and $\lambda=2L$ for fundamental mode). In thin tube (TT) approximation, the phase speed $V_{ph}$ is {the} same as the kink speed $C_K$ \citep{nv} :
\begin{equation}
    V_{ph}=C_K=\frac{2L}{P} ,
\label{vph}
\end{equation}  
where $C_K$ is also defined as the density averaged Alfv\'en speed \citep{er}, {\it i.e} , 
\begin{equation}
C_K=\sqrt{\frac{\rho_0C_{A0}^2+\rho_eC_{Ae}^2}{\rho_0+\rho_e}}~,
\label{ck}
\end{equation}
{where} $\rho_{0}$ and $\rho_{e}$ are the density of uniform plasma inside and outside the loop respectively and $C_{A0}$ and $C_{Ae}$ are the internal and external Alfv\'en speeds which can be defined as follows :\\
\begin{equation}
C_{A0}=B_0/\sqrt{\mu_0 \rho_0 }~,
\label{ca0}
\end{equation}
\begin{equation}
C_{Ae}=B_e/\sqrt{\mu_0 \rho_e }~,
\label{cae}
\end{equation}
{where} $B_{0}$ and $B_{e}$ are the internal and external magnetic fields respectively for the coronal loop.

Substituting {Equation (\ref{ca0}) and (\ref{cae}) into Equation (\ref{ck})}, and assuming that the internal and external magnetic fields of the loop are equal, we get :
\begin{equation}
C_{A0}=\frac{C_K}{\sqrt{\frac{2}{1+\frac{\rho _e}{\rho_0}}}}~,
\label{ck3}
\end{equation}
Using {Equation (\ref{vph})}, we estimate the kink speed ($C_K$) to be $\sim$ $392\pm17$~km~s$^{-1}$ (and $\sim$ $560\pm38$~km~s$^{-1}$, assuming coronal loop to be {semicircle}) and the internal Alfv\'en velocity $\sim$ $299\pm27$~km~s$^{-1}$ (and $\sim$ $415\pm28$~km~s$^{-1}$, assuming coronal loop to be a semicircle), assuming density contrast $\left(\frac{\rho _e}{\rho_0}\right)$ = 0.1.
It is worth noting at this point that density contrast depends on the EUV intensity ratio as,  $\left(\frac{\rho _e}{\rho_0}\right)^{2}=\frac{I_{e}}{I_{o}}${, where} $I_{e}$ and $I_{o}$ are EUV intensity of background and inside the loop, respectively. We find that the intensity ratio along the loop changes from loop footpoint to the loop top. Furthermore, we also note that the intensity ratio at a given position along coronal loop changes as it oscillates, presumably due to variations in the column depth of the loop along the line-of-sight by the wave \citep{cooper2003}. We estimate the intensity ratio at several places along the loop and at different times. We find that the intensity ratio varies from 0.18 to 0.5. Thus density contrast varies from 0.4 to 0.7. We use AIA 171~\AA~for the analysis because the loop is best visible in this wavelength. However, it is important to note that the intensity contrast of the loop also depends on its orientation thus the density {contrast} may still be a rough estimate. Table~\ref{table1} shows the estimation of Alfv\'en speed and magnetic field strength for different values of density contrast ({\it i.e,} 0.1, 0.4 and 0.7). We keep 0.1 for comparison since this is a typical value which is used in most studies.
Using {Equation (\ref{ca0})} and taking $\mu_0=4\pi\times10^{-7}$~H~m$^{-1}$ (in SI units), we can estimate the magnetic field inside the coronal loop ($B_0$),  provided the internal density of the coronal loop is known.

\begin{table}
\caption{Estimation of Alfv\'en speed and magnetic field strength inside the coronal loop.}
\label{table1}
\begin{tabular}{lccc|ccc}     % define the column alignment
                           % l: left, c: center, r: right
\hline
\multicolumn{1}{c}{ } & \multicolumn{3}{c|}{Projected length} & \multicolumn{3}{c}{{Semicircular} model} \\
  \hline                   % horizontal line
$\frac{\rho _e}{\rho_0}$   & 0.1 & 0.4 & 0.7 & 0.1 & 0.4 & 0.7 \\
  \hline
$C_{A0}$(km~s$^{-1}$) &  $299\pm27$ & $328\pm14$ & $361\pm16$ & $415\pm28$ & $469\pm32$ & $516\pm35$\\
$B~(\rm{G})$ & 2.68$\pm$0.64 &  2.86$\pm$0.12 &  3.20$\pm$0.14 &  3.62$\pm$0.24 &  4.1$\pm$0.28 &  4.5$\pm$0.31\\

  \hline
\end{tabular}
\end{table}

\subsection{Density estimate using DEM analysis}\label{secdem}
\begin{figure}[h!]
  \begin{center}
    \includegraphics[scale=0.5]{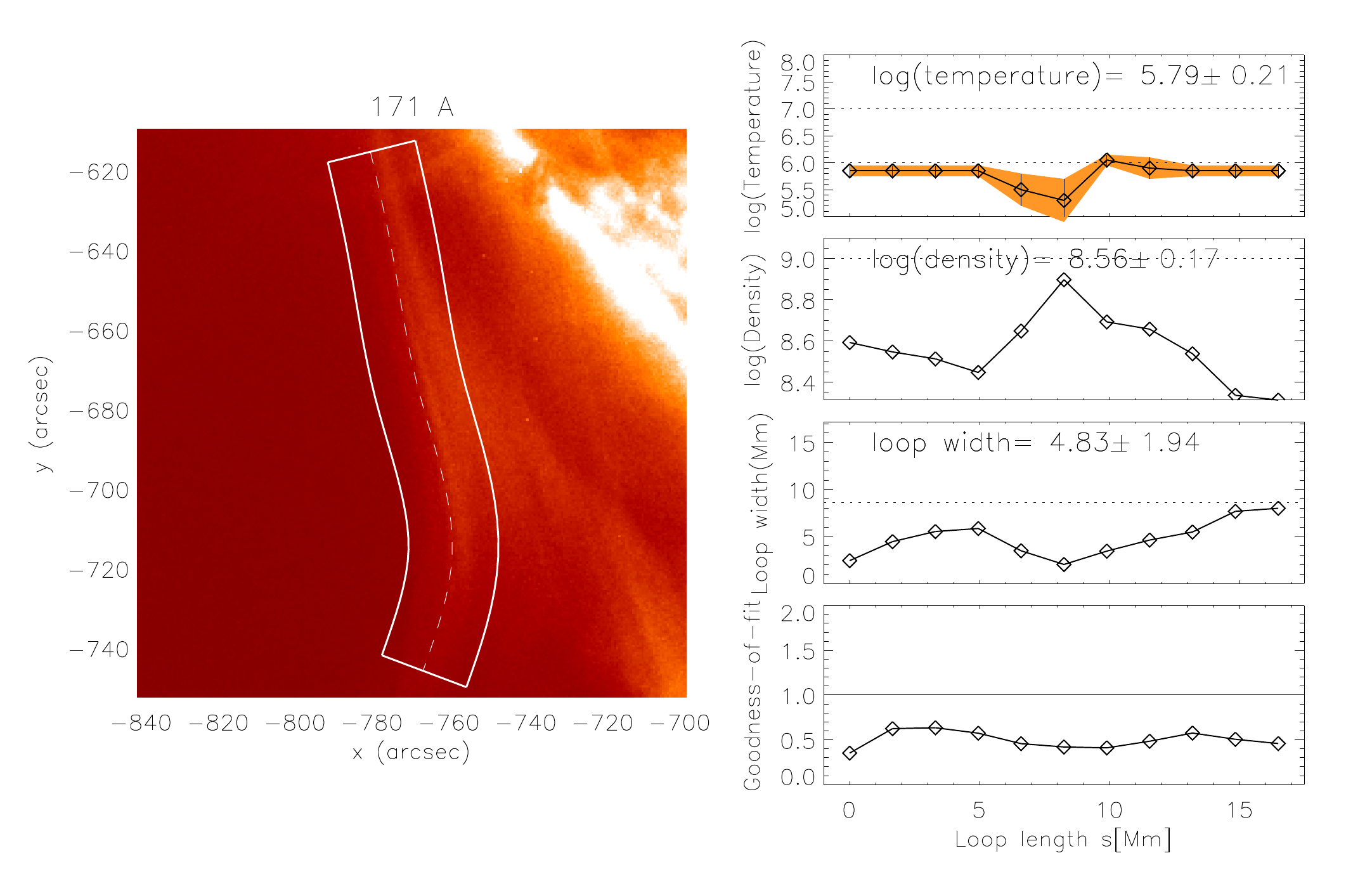}
    \caption{{The left} panel shows the location of loop, enclosed in curves, selected for analysis. The best-fit values of the DEM peak temperature, the electron densities, loop widths, and the goodness-of-fit $\chi^2$ for the 171~\AA, Gaussian DEM fits, are shown in the right panel of the graph.}

    \label{dem}
  \end{center}
\end{figure}

In order to calculate the internal density of coronal loop we use differential emission measure analysis technique developed by \inlinecite{aschdem}. 
%Using this, first we generated the temperature and emission maps for the loop. This is shown in {Figure}~\ref{teem}.
%\begin{figure}[h!!!!!!!!!!!!!!!!!!!!!!!!!!!!]
 % \begin{center}
%    \includegraphics[height=6.5cm,width=11.5cm]{aiaa_teem_map.eps}
 %     \caption{Temperature and emission measure maps of the coronal loop}
  %   \label{teem}
 % \end{center}
%\end{figure}
Using this automated technique we obtain the number density of electrons ($n_e$) to be $10^{8.56\pm0.17}$~cm$^{-3}$ and the average temperature to be $10^{5.79\pm0.21}$~K (see {Figure}~\ref{dem}).

%%%%%%%%%%%%%%%%%%%%%%%%%%%%%%%%%%%%%%%%%%%%%
%%%%%%%%%%%%%%%%%%%%%%%%%%%%%%%%%%%%%%%%%%%%%
%%%%%%%%%%%%%%%%%%%%%%%%%%%%%%%%%%%%%%%%%%%%%
%%%%%%%%%%%%%%%%%%%%%%%%%%%%%%%%%%%%%%%%%%%%%

\subsection{Calculation of magnetic field inside the coronal loop}\label{calc}
Estimating electron density using DEM and assuming H:He=10:1, we obtain the magnetic field strength inside the coronal loop for different values of density contrast using {Equation (\ref{ca0}) and (\ref{ck3}) (see Table~\ref{table1})}. The estimated value of magnetic field inside the loop is low as compared to active region loops.\\

\section{Energy estimates}\label{energy}
In order to understand if the jet can trigger the transverse oscillations in the coronal loop, we calculate the energy density of the jet and compare it with the energy density of the oscillating coronal loop. Although both transverse oscillations and linear motion {contribute} to the energy density of the jet, we estimate energy density of the jet due to only linear motion. Moreover, we estimate energy density using the bright feature of the jet. We do not consider the energy density of hot component of jet since it is very difficult to measure its density and thus its energy density. Therefore, it should be borne in mind that the energy density of jet is an underestimate of total energy density. 
The energy density of the propagating jet due to its linear motion is defined {as}
\begin{equation}
E_{jet}=\frac{1}{2}\rho_{jet} v^2~,
\end{equation}
{where} $\rho_{jet}$ is the density of the jet. 
%We use typical value of density of jet since it is hard to estimate the density of jet with DEM as it becomes very faint and hardly visible in EUV channels. 
The typical value of the {number density of the jet} is reported between $10^9$~cm$^{-3}$ \citep{shimojo2000} to $10^{10}$~cm$^{-3}$, \citep{madjarska2011,insertis2013}.\\
We also estimate the number density of jet using DEM at the instant when it is clearly observed. The density is estimated to be $10^{9.92}$~cm$^{-3}$.
$v$ is the velocity of the jet which is found to be $\sim43\pm4$~km~s$^{-1}$ (see {Section}~\ref{sec4}).
Using {$n_{jet}=10^{10}$~cm$^{-3}$} and velocity of jet, we {estimate} the energy density to be $(19.6\pm3.6)\times 10^{-3}$~J~m$^{-3}$.

Energy density of the oscillating loop is defined as, \citep{goosens2013}:
\begin{equation}
E_{monolithic}=\frac{1}{4}\rho_0 \omega^2 {A_0}^2~,
\label{mono}
\end{equation}
{where} $\rho_0$ is the internal density of the coronal loop. {Using DEM we estimate the number density inside the loop to be $10^{8.56\pm0.17}$ cm$^{-3}$}. $\omega$ is the angular frequency of the oscillating loop ($\omega=kC_K=\frac{\pi}{L}C_K$, where, $k=\frac{2\pi}{\lambda}$ and $\lambda=2L$ for fundamental mode of vibration). Thus, $\omega=(3.36\pm0.36)\times10^{-3}$ s$^{-1}$ (and 3.27$\pm$0.19$\times$10$^{-3} s^{-1}$ assuming loop to be {semicircle}) using the parameters estimated in {Section}~\ref{calc}. $A_0$ is the displacement amplitude.

{The displacement amplitude for n=4 slice is $\sim4.50\pm1.72$ Mm} which is maximum of all the slices (see {Figure}~\ref{fitting}). We estimate the energy density of the loop corresponding to this amplitude is $\sim(3.28\pm0.89)\times10^{-5}$J~m$^{-3}$ ($\sim(3.26\pm0.23)\times10^{-5}$J~m$^{-3}$ assuming loop to be a {semicircle}). 
For n=7 slice, where the displacement amplitude is $\sim2.75\pm0.22$~Mm which is minimum of all the slices, the energy density is estimated to be $\sim(1.23\pm0.09)\times10^{-5}$~J~m$^{-3}$ ($\sim(1.22\pm0.1)\times10^{-5}$~J~m$^{-3}$ assuming {the loop} to be a {semicircle}). Therefore, the {energy density of the oscillating loop is within the range from $(1.22-3.28)\times10^{-5}$~J~m$^{-3}$}.\\
Since the coronal loop is not monolithic and consists of many fine loops, we estimate the energy density using multistranded loop model. The energy density in multistranded oscillating coronal loop is {given by}
\begin{equation}
E_{multistrand}=\frac{1}{2} f (\rho_0 + \rho_e) \omega^2 {A_0}^2~,
\label{multi}
\end{equation} 
 {where} $\rho_e$ is the external density of the coronal loop. $f$ is the filling factor, which is defined as the ratio of the sum of volume of individual flux tube ensemble together to the total volume containing flux tubes. Using {Equation (\ref{mono}) and (\ref{multi})} we get
\begin{equation}
\frac{E_{multistrand}}{E_{monolithic}}=2 f (1+\frac{\rho_e}{\rho_o} )~,
\label{multibymono}
\end{equation} 

We calculate the filling factor, $f$ as defined above at several instances along the length of coronal loop. The mean and standard deviation is estimated. Therefore, the $f$ is found out to be 0.17 $\pm$ 0.03. Substituting the value of {\textit{f}} in {Equation~(\ref{multibymono})}, we find that $E_{multistrand}=0.434$ $E_{monolithic}$ for density constrast of 0.1. $E_{multistrand}=0.476$ $E_{monolithic}$ and  $E_{multistrand}=0.578$ $E_{monolithic}$ assuming density contrast to be 0.4 and 0.7, respectively.
Therefore, we find that the energy density of the jet is much greater than the energy density of the coronal loop. The jet might be transferring a part of its energy to displace the loop. Hence, {the jet} can be inferred to be the cause of the transverse oscillations in the coronal loop.

\section{Interaction between jet and loop}\label{sec7}
In the {Section}~\ref{energy}, we find that the jet {has} enough energy to excite oscillations in the coronal loop. In running difference movies of AIA 211~\AA~(see movie~7) we find the evidence of both hot and cool {components} of a jet hitting the coronal loop. This could be a simple projection effect, but we do not know which trajectory the jet follows: maybe it curves, {hits} the loops and some part of the jet plasma is seen as the inward moving feature in STEREO/EUVI~195~\AA.

Furthermore, there can also be the possibility of the collective transverse oscillation in two fluxtubes. Looking the jet's magnetic field spine constitutes a fluxtube that {includes} flowing plasma {and is subject to} long-period (24 min) transverse oscillations. The coronal loop is another fluxtube {originally in equilibrium} in the vicinity of jet's plasma column. The separation between the oscillating jet and coronal loop is estimated to be $\sim 45$ Mm as shown in {Figure}~\ref{jet_s} (right panel).
%%%%%%%%%%%%%%%%%%%%%%%%%%%%%%%%%%%%%%%%%%%%%
%%%%%%%%%%%%%%%%%%%%%%%%%%%%%%%%%%%%%%%%%%%%%
%\begin{figure}
  %\begin{center}
    %\includegraphics[scale=0.3]{fig9.eps}
   %\caption{Figure showing the separation between the oscillating jet and coronal loop.}
  % \label{jet_distance}
   %  \end{center}
 %    \end{figure}
     %%%%%%%%%%%%%%%%%%%%%%%%%%%%%%%%%%%%%%%%%%%%%
%%%%%%%%%%%%%%%%%%%%%%%%%%%%%%%%%%%%%%%%%%%%%
\inlinecite{2008ApJ...676..717L} studied the collective kink oscillations of two identical parallel magnetic fluxtubes.
They found that there are four modes of oscillations of {such a} system in which two are in-phase oscillations.
Later, \inlinecite{2008A&A...485..849V} reported that the kink
oscillations of {the system of} two non-identical {tubes is} degenerate. They found that {similarly} to the
kink oscillations of a single tube {with} circular cross-section, there {is} no preferable direction of kink oscillation polarization, and the two long-period (and also short-period) oscillations merge with each other forming two oscillatory modes (one with long and another short period one).

Significant developments have also been made considering the various models demonstrating the 
damping of the kink oscillations {in} multiple magnetic fluxtubes \citep{2009ApJ...692.1582L,2010ApJ...716.1371L,2009ApJ...694..502O,2008ApJ...679.1611T}. The linear theory of the resonant damping of kink oscillations in two parallel magnetic tubes is also developed by \inlinecite{robertson11}. It should be noted {that the} observed jet shows {transverse} kink oscillations of two periods : 21 min (long) and 2 min (short) ({Figure}~\ref{jet_xt}). The {loop} in the vicinity also {exhibits} similar transverse oscillations, although only long period (31 min) is detectable in the observational base line. It is also noticeable that the long period kink waves in both {flux systems} are almost out of phase. It should be borne in mind that we do not know that at what location and at what time the jet interacts with the loop. It may not be necessary that the visible jet material has to strike the loop. Both {the} jet and coronal loop are magnetic flux tubes thus the interaction between them may happen earlier than {the observational signatures it produces}. If {the jet} interacted with the loop and triggered oscillations before the bright visible material {reached} the same location, in that scenario, at the instant when bright material will appear adjacent to coronal loop both {systems} might go out of phase. {Both} fluxtubes, \textit{i.e.}, upper magnetoplasma column of {the} jet and the part of coronal loop in its vicinity (45 Mm apart), are non-identical fluxtubes as their plasma and magnetic field properties are essentially not the same. It is worth noting at this point that this distance is projected in the plane of sky and therefore is an underestimate of the real distance which could be higher. Therefore, the model of \inlinecite{2008A&A...485..849V} might be at work in the present observational base-line where the transverse oscillations of short period (2.0 min) and long period (21 min) both are excited in the jet ({Figure}~\ref{jet_xt}). The two modes with higher period are most likely merged in one degenerate mode, and the same occurred with the two modes with the lower period. Since
the modes from each pair are polarized in the mutually orthogonal directions (cf., {Figure}~\ref{jet_xt}), the degenerate modes created by merging of two
modes can have arbitrary polarization. The long-period mode is only detected in the coronal loop tube, while the short period is not detected. This may be the fact that plane-of-loop apex is in such an orientation that the arbitrarily set polarization direction of the short-period oscillations (what is seen in the jet's body) is non-identifiable.

\section{Conclusions}
We report large amplitude and long period transverse oscillations in a coronal loop most likely triggered by a nearby oscillating jet. We find that the jet appears as bright feature in all AIA channels but soon disappears {from hot channels} like 211~\AA. A {faint} emission is also seen in AIA 211~\AA~which is {considered as a} signature of heating. We estimate the length of the loop by two different methods. (i) Calculating the projected distance of the loop in the plane of sky, which can be considered as the lower limit of the length of the loop and (ii) estimating the length of the loop assuming a {semicircular} geometry, which can be considered as the upper limit of the length of the loop. Therefore we estimate the lower (upper) limit of kink speed and Alfv\'en speed to be $\sim 392\pm17$ ($560 \pm 38$)~km~s$^{-1}$ and $\sim299\pm27$ ($516 \pm 35$)~km~s$^{-1}$. We {use the seismic} inversion technique to estimate {the magnetic field strength} inside the coronal loop. We estimate the lower (upper) limit of the strength of the magnetic field inside coronal loop to be $\sim2.68\pm0.64$ ($4.5\pm 0.31$) G which is less than a typical value of magnetic field inside a coronal loop. It might be because it is not an active region coronal loop. Finally we estimate the energy density of loop. We find that the energy density varies from $1.22\times10^{-5}$~J~m$^{-3}$ to 3.28$\times10^{-5}$~J~m$^{-3}$ which is about two to three order of magnitude lower than the energy density of the jet ($\sim(19.6\pm3.6)\times 10^{-3}$~J~m$^{-3}$). Therefore, we conclude that the energy stored in the jet is found to be enough to excite oscillations in the coronal loop. The present observations also support the model of the collective transverse oscillations of two non-identical magnetic fluxtubes in which long and short period modes are excited with arbitrary direction of polarization \citep{2008A&A...485..849V}

\section{Acknowledgements}
We thank Dr. M. Ruderman for fruitful discussion. We also thank anonymous {referee} for her/his useful suggestions which has {enabled} us to improve the content of manuscript substantially.

%\bibliographystyle{spr-mp-sola}
%\bibliography{report_solarphysics_nov_2016.bib}

\end{article} 
\end{document}